# The Role of Explainable AI in Revolutionizing Human Health Monitoring: A Review

Abdullah Alharthi [1] Ahmed Alqurashi [2], Turki Alharbi[3], Mohammed Alammar [4], Nasser Aldosari [5], Houssem Bouchekara [6], Yusuf Shaaban [7], Mohammad Shoaib Shahriar [8], and Abdu lrahman Al Ayidh [9]

[1] Department of Electrical Engineering, King Khalid University, Abha, Saudi Arabia, asalharthi@kku.edu.sa
[2] Department of Electrical Engineering, Ummul Al Qura University Makkah, Saudi Arabia, akqurashi@uqu.edu.sa
[3] Department of Electrical Engineering, Taif University, Taif, Saudi Arabia, turki.alharbi@tu.edu.sa
[4] Department of Electrical Engineering, King Khalid University, Abha, Saudi Arabia, mmalamar@kku.edu.sa
[5] King Abdullah Medical City, Makkah, Saudi Arabia, aldosari.n@kamc.med.sa
[6] Department of Electrical Engineering, University of Hafr Al Batin, Hafr Al Batin, Saudi Arabia, rbouchekara@uhb.edu.sa
[7] Department of Electrical Engineering, University of Hafr Al Batin, Hafr Al Batin, Saudi Arabia, shaaban@uhb.edu.sa
[8] Department of Electrical Engineering, University of Hafr Al Batin, Hafr Al Batin, Saudi Arabia, mshoaib@uhb.edu.sa
[9] Department of Electrical Engineering, King Khalid University, Abha, Saudi Arabia, aalaid@kku.edu.sa

corresponding authors, Abdullah Alharthi: asalharthi@kku.edu.sa & Yusuf Shaaban: shaaban@uhb.edu.sa

**Abstract:** The complex nature of disease mechanisms and the variability of patient symptoms pose significant challenges in developing effective diagnostic tools. Although machine learning (ML) has made substantial advances in medical diagnosis, the decision-making processes of these models often lack transparency, potentially jeopardizing patient outcomes. This review aims to highlight the role of Explainable AI (XAI) in addressing the interpretability issues of ML models in healthcare, with a focus on chronic conditions such as Parkinson's, stroke, depression, cancer, heart disease, and Alzheimer's disease. A comprehensive literature search was conducted across multiple databases to identify studies that applied XAI techniques in healthcare. The search focused on XAI algorithms used in diagnosing and monitoring chronic diseases. The review identified the application of nine trending XAI algorithms, each evaluated for their advantages and limitations in various healthcare contexts. The findings underscore the importance of transparency in ML models, which is crucial for improving trust and outcomes in clinical practice. While XAI provides significant potential to bridge the gap between complex ML models and clinical practice, challenges such as scalability, validation, and clinician acceptance remain. The review also highlights areas requiring further research, particularly in integrating XAI into healthcare systems. The study concludes that XAI methods offer a promising path forward for enhancing human health monitoring and patient care, though significant challenges must be addressed to fully realize their potential in clinical settings.

## 1. Introduction

Artificial Intelligence (AI) has not just significantly impacted healthcare in recent years, but it has also revolutionized how we monitor and manage human health. AI has opened new avenues for disease prevention, diagnosis, management, and treatment [1]. It has addressed health issues in various ways, driving advancements in healthcare delivery. These advancements must align with Healthcare Management System (HMS) standards, highlighting the need for medical devices that are cost-efficient, environmentally friendly, durable, and accurate [2].

AI is essential in health monitoring and managing large volumes of data, such as patient medical records and real-time biometrics. This capability is crucial for creating predictive models of disease progression, personalizing treatment plans, and enhancing remote patient monitoring systems. AI's potential ranges from fitness trackers that monitor daily activity to advanced systems predicting chronic disease onset and progression [3].

AI has become increasingly integral to healthcare with advancements in Machine Learning (ML), Deep Learning (DL), and Convolutional Neural Networks (CNNs). These AI models are transforming medical practices, particularly in early diagnosis and management of chronic diseases like cardiovascular conditions, cancer, and diabetes [4], [5]. AI-driven technologies, such as advanced robotic surgeries, help reduce misdiagnosis and unnecessary hospitalizations, saving time and money [6]. Additionally, AI, especially through DL models like Residual Networks (ResNet), has shown remarkable success in diagnosing complex conditions such as Alzheimer's disease and various stages of lung cancer using techniques like Resting-State Functional Magnetic Resonance Imaging (rs-fMRI) [7]. These advancements highlight AI's potential to enhance the accuracy and efficiency of healthcare interventions.

The evolution of AI in healthcare represents a significant journey marked by substantial advancements and paradigm shifts. Initially, the role of AI in healthcare was predominantly focused on data management and basic pattern recognition. This included tasks like managing patient records and assisting in diagnostic procedures by identifying patterns in medical images [1]. These early implementations of AI, while revolutionary for their time, were mostly rule-based systems and lacked the sophistication seen in later developments.

As AI technology progressed, the introduction of more advanced ML algorithms and DL brought a significant expansion in its capabilities within the healthcare sector. AI systems started to take on more complex tasks such as predictive analytics for patient outcomes, personalized treatment recommendations, and assistance in surgical procedures [8]. However, this advancement in AI technology highlighted a major challenge: the problem of interpretability.

Explainable AI (XAI) emerged to address this challenge. XAI aims to make the decision-making processes of AI systems clear and understandable, which is particularly important in healthcare, where these decisions can greatly affect patient lives [9]. It is essential for both healthcare professionals and patients to understand and justify the reasoning behind a medical recommendation or diagnosis made by an AI system. By explaining how AI models operate, XAI builds trust and supports better decision-making in healthcare settings [10].

The move towards using XAI in healthcare is an important development in the field. It improves the reliability of AI applications and ensures they meet ethical standards, making AI-driven healthcare solutions not only effective but also accountable and understandable [11]. Integrating XAI into healthcare AI systems is an evolving area, with ongoing research focused on improving the clarity and usefulness of the explanations provided by these systems [11,12]. This shift represents a commitment to more responsible and patient-focused applications of AI in healthcare, ensuring that technology aids in diagnosis and treatment in a way that is clear and understandable. XAI's main goal is to make the operations and decisions of AI models transparent, understandable, and interpretable to human users [13]. This goal involves several key principles, including transparency, interpretability, and accountability, all of which are crucial for building trust and ensuring effective collaboration between humans and AI [14].

The development of XAI is closely related to the limitations of earlier AI models. As AI systems grew more complex, especially with the rise of DL and Artificial Neural Networks (ANNs), their decision-making processes became harder to understand. This "black box" nature of AI created significant challenges in fields where understanding the reasoning behind decisions is critical, such as healthcare, finance, and legal systems [15,16]. XAI was created to address this issue, with the goal of making AI processes clearer and more trustworthy for users, regardless of their technical background [17]. This push towards explainability in AI is not just a technical upgrade; it represents a fundamental shift in how AI is perceived and utilized across various sectors. It acknowledges the need for AI systems to be robust, efficient, transparent, and understandable, ensuring they can be effectively integrated into societal structures and processes [18].

This review addresses a critical gap in human health monitoring by focusing on the role of Explainable AI (XAI) in healthcare, an area often overlooked in previous studies. While earlier reviews have primarily emphasized advancements in AI for predictive modeling, diagnostics, and personalized medicine [19]-[21], they frequently neglect the importance of explainability. Understanding AI model decisions is crucial in healthcare, as these decisions directly impact patient outcomes. Although many studies concentrate on AI's performance in disease diagnosis and risk prediction, they rarely explore how to make these models interpretable or how XAI can enhance trust and reliability in clinical settings [22]. This review underscores the significance of integrating XAI into healthcare prediction models, advocating for AI systems that are both effective and transparent. It aims to guide researchers and practitioners in selecting the most suitable machine learning (ML) and deep learning (DL) methods for rapid and reliable disease detection and classification.

By providing a comprehensive review of published work on AI, with a specific focus on XAI methods in healthcare, this paper contributes to the field by highlighting trending XAI algorithms and their applications. It assesses their suitability for various diseases, offering valuable insights for future research. What sets this review apart is its emphasis on the practical implementation of XAI methodologies in patient health monitoring and their integration into clinical workflows. Unlike prior reviews that offer broad discussions on XAI applications in healthcare, this work prioritizes the real-world contributions of explainable models to patient care. This focus underscores the review's relevance in advancing XAI from research to clinical practice, bridging the gap between theoretical advancements and practical healthcare solutions.

The paper is organized as follows, we begin by summarizing the different ML and XAI methods used in diagnosing various diseases, as well as the literature search strategy. Section 3 presents the findings from our literature search, focusing on XAI techniques and their application to specific diseases or diagnostic features. Section 4 discusses the results, anticipated trends, and challenges. Section 5 concludes with a summary of the key points.

## 2. Basics and Background

This section presents an overview of the widely used ML techniques for health monitoring and disease detection. Additionally, an overview of the commonly used XAI methods that have been applied in the field of human health monitoring is provided here. Further we describe the literature search strategy.

### 2.1. Machine learning (ML)

ML is a field of AI that enables computers to learn and improve from experience without being explicitly programmed. It involves developing algorithms and statistical models that allow systems to perform specific tasks effectively without relying on rule-based programming. ML algorithms build a mathematical model based on sample data, known as "training data", to make predictions or decisions without being directly programmed. This allows machines to adapt and improve their performance over time as they are exposed to more data.

Most of the wildly used ML algorithms are Supervised Learning (SL) as shown in Figure 1. It includes Linear Regression (LR) for predicting continuous outputs, Logistic Regression for binary classification, Decision Trees for modeling decisions, Random Forests (RF) for ensemble learning, Support Vector Machine (SVM) for optimal separation, and K-Nearest Neighbors (KNN) for proximity-based classification. Unsupervised Learning features K-Means Clustering for partitioning data, Hierarchical Clustering for hierarchical analysis, Principal Component Analysis for dimensionality reduction, Autoencoders for feature extraction, and Gaussian Mixture Models for probabilistic clustering. Reinforcement Learning involves Q-Learning, Deep Q-Networks, Policy Gradient Methods, and Actor-Critic Methods for optimizing reward-based actions. DL covers ANNs, CNNs for image processing, Recurrent Neural Networks (RNNs) for sequential data, Long Short-Term Memory for long-term dependencies, Transformers for sequential data and natural language processing, Generative Adversarial Networks (GANs) for generating synthetic data, Graph Neural Networks (GNNs)for graph-structured data, and Federated Learning for distributed data across devices. ML finds applications in computer vision, natural language processing, robotics, and predictive analytics, with algorithm choice dependent on the specific problem, data type, and desired outcome.

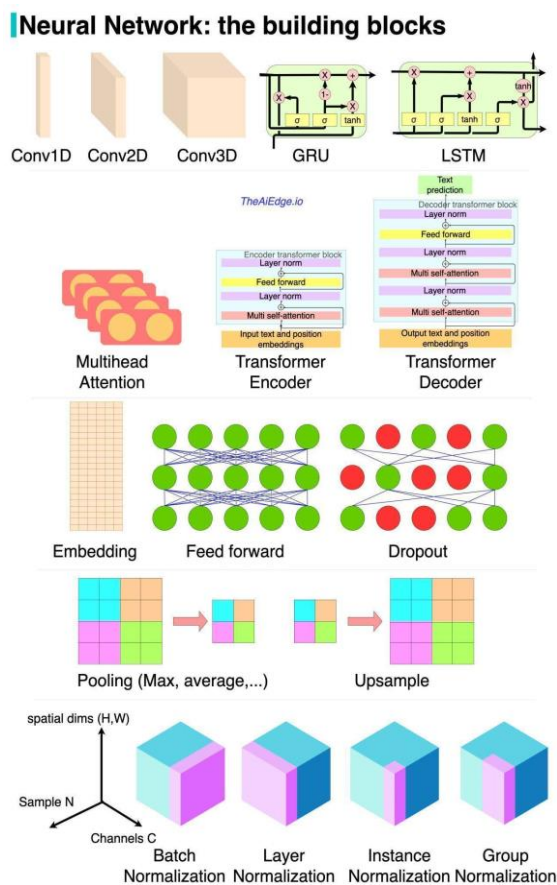

**Figure 1.** ML methods: CNNs, LSTM, transformers ANNs with feature maps, and layer types.

## 2.2. Accuracy measures for ML models

The widely used accuracy metrics for top ML algorithms are: Mean Squared Error (MSE), R-squared (coefficient of determination), Accuracy- Precision- Recall- F1-score- Area Under the Receiver Operating Characteristic (ROC) Curve (AUC-ROC).

- **Mean Squared Error (MSE)** measures the average squared difference between actual and predicted values, with lower values indicating a better fit [23]. It is particularly sensitive to outliers and is commonly used in regression problems. With $y_i$ is the actual value and $\hat{y}_i$ is the predicted value, the formula for $MSE$ [23] is:

$$MES = \frac{1}{n} \sum_i^n (y_i - \hat{y}_i)^2, \quad (1)$$

- **R-squared (Coefficient of Determination) represents** the proportion of variance in the dependent variable that is predictable from the independent variables [24]. Values range from 0 to 1, with higher values indicating a better fit. It is also used in regression problems. With $\underline{y}_i$ is the mean of the actual values the predicted value the formula for R-squared [24] is:

$$R^2 = 1 - \frac{\sum_i^n (y_i - \hat{y}_i)^2}{\sum_i^n (y_i - \hat{y}_i)^2}, \quad (2)$$

- **Accuracy** is the ratio of correctly predicted instances to **the** total instances and is suitable for balanced datasets [25]. It is commonly used in classification problems. With TP is the number of true positives, TN is the number of true negatives, FP is the number of false positives, FN is the number of false negatives. The formula for accuracy [25] is:

$$Accuracy = \frac{TP + TN}{TF + TN + FF + FN}, \quad (3)$$

- **Precision** is the ratio of true positive predictions to the sum of true positive and false positive predictions, making it useful when the cost of false positives is high. It is particularly relevant in binary classification problems. The formula for $precision$ [25] is:

$$Precision = \frac{TP}{TP+FP}, \quad (4)$$

- **Recall (Sensitivity)** is the ratio of true positive predictions to the sum of true positive and false negative predictions, making it useful when the cost of false negatives is high. It is also particularly relevant in binary classification problems. The formula for $recall$ [25] is:

$$Recall = \frac{TP}{TP+FN}, \quad (5)$$

- **F1-score** is the harmonic means of precision and recall, providing a single metric that balances both concerns. It is useful when the dataset is imbalanced and is commonly used in binary classification problems. The formula for the $F1-score$ [25] is:

$$F1 - score = 2 \times \frac{Precision \times Recall}{Precision + Recall}, \quad (6)$$

- **Area Under the Receiver Operating Characteristic Curve (AUC-ROC)** measures the ability of the model to distinguish between classes [26]. The higher the AUC, the better the model at predicting 0s as 0s and 1s as 1s. It is used in both binary and multi-class classification problems. The AUC-ROC is the area under the ROC curve, which plots TPR (True Positive Rate) against FPR (False Positive Rate) at various threshold settings. With $I_P$ and $I_n$ denotes positive and negative data samples and $R_i$ is the rating of the $ith$ positive samples. the formula for $AUC$ [26] is:

$$AUC = \frac{\sum_0^n R_i (I_P - I_P(\frac{I_P+1}{2}))}{I_P + I_n}, \quad (7)$$

## 2.3. Explainable Artificial Intelligence (XAI)

XAI refers to AI systems and techniques that are designed to provide clear, understandable, and interpretable explanations of how they make decisions and predictions. As aforesaid. The primary goal of XAI is to make AI systems more transparent, trustworthy, and accountable by enabling humans to comprehend and trust their outputs. This is crucial in high-stakes fields like healthcare, finance, and autonomous driving, where understanding the reasoning behind AI decisions is essential for safety, compliance, and ethical considerations.

Key aspects of XAI include:
1. Transparency: The ability of the AI model to provide insights into its inner workings and decision-making processes.
2. Interpretability: The ease with which a human can understand the reasoning behind an AI model's output.
3. Accountability: Ensuring that AI systems can be held responsible for their decisions and that these decisions can be audited.
4. Trust: Building confidence in AI systems by providing clear and comprehensible explanations of their behavior.

Here is a list of some of the most widely used algorithms and techniques in XAI in the following subsections.

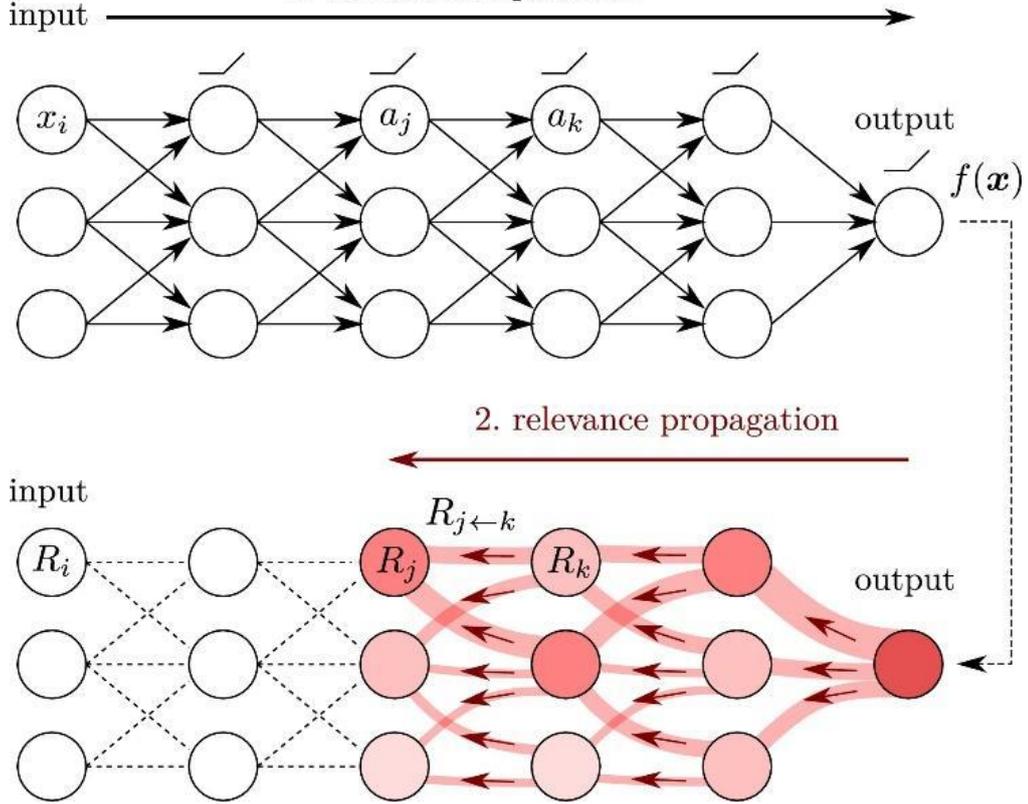

**Figure 2.** The graph illustrates LRP, which is a method that identifies important pixels by running a backward pass in the neural network. The backward pass is a conservative relevance redistribution procedure, where neurons that contribute the most to the higher layer receive most relevance from it [28].

### 2.3.1 Layer-wise Relevance Propagation (LRP)

Layer-wise Relevance Propagation (LRP) (see Figure 2) is primarily used for explaining DL models, particularly ANNs [27-31]. It works by propagating the prediction score backward through the network to assign relevance scores to each input feature. The process begins with the final output of the ANN, which is the prediction score. This score is then distributed backward through the network layers using specific propagation rules designed to maintain relevance conservation, ensuring that the total relevance at each layer matches the prediction score. At each layer, the relevance score assigned to each neuron is divided among its predecessors based on their contributions to the neuron's activation. This process continues until the input layer is reached. The output of LRP is a set of relevance scores for each input feature, indicating their contribution to the final prediction and providing an interpretable mapping from input features to the model's decision.

We first note that an ANN consists of multiple layers of neurons, where neurons are activated as follows [32]:

$$a_k = \sigma(\sum_j^n a_j \omega_{jk} + b_k), \tag{8}$$

Here, $a_k$ the neuron activation and $a_j$ is the activation of the neuron in the previous layer in forward direction; $\omega_{jk}$ denote the weight received in forward direction by neuron $k$ from neuron $j$ in the previous layer and $b_k$ is the bias. The sum is computed over all the $j^{th}$ neurons that are connected to $k^{th}$ neuron. $\sigma$ is a nonlinear monotonically increasing activation function. These activations, weights, and biases are learned by the DL model during supervisory training. During training, the output $f_c(x)$ is evaluated in a forward pass and the parameters $(\omega_{jk} + b_k)$ are updated by back-propagating using model error.

The LRP approach decomposes the DL output for a given prediction function of multi class $c$ as $f_c$ for input $x_i$ and generates a "relevance score" $R$ for the $i^{th}$ neuron received from $R_j$ for the $j^{th}$ neuron in the previous layer which is received from $R_k$, for the $k^{th}$ neuron in the lower layer, where the relevance conservation principle is satisfied [31] as:

$$\sum_i^n R_{i \leftarrow j} = \sum_j^n R_{j \leftarrow k} = \sum_k^n R_k = f_c(x), \tag{9}$$

There are other propagation rules such as ($\alpha\beta$-rule) [219]:

$$R_j = \sum_k^n (\alpha \frac{a_j \omega_{jk}^+}{\sum_j^n a_j \omega_{jk}^+} - \beta \frac{a_j \omega_{jk}^-}{\sum_j^n a_j \omega_{jk}^-}) R_k, \tag{10}$$

Where each sum corresponds to $R_{j \leftarrow k}$ a relevance message and $a_j \omega_{jk}^+$ and $a_j \omega_{jk}^-$ denote the positive and negative part of $a_j \omega_{jk}$ respectively. The parameters $\alpha$ and $\beta$ are chosen so that $\alpha - \beta = 1$ and $\beta \geq 0$. A propagation rule can be chosen by selecting $\beta = 0$ to in result the following rule:

$$R_j = \sum_k^n \frac{a_j \omega_{jk}^+}{\sum_j^n a_j \omega_{jk}^+} R_k, \tag{11}$$

There are other stabilizing terms that can be used to avoid divisions by zero as explained in [27], [28]. For the LRP-$\gamma$ rule, let the neurons interconnection be as follow:

$$a_k = max(0, \sum_{0,j}^n a_j \omega_{jk}), \tag{12}$$

Here, $a_j$ denote input activation and $\omega_{jk}$ denote the weight received by neuron $j$ from neuron $k$ in the above layer. The sum is computed over all neurons $j$ in the lower layer plus a bias term $\omega_{0k}$ with $a_0 = 1$. The LRP-$\gamma$ rule is given by:

$$R_j = \sum_k^n \frac{a_j(\omega_{jk} + \gamma \omega_{jk}^+)}{\sum_{0,j}^n a_j(\omega_{jk} + \gamma \omega_{jk}^+)} R_k, \tag{13}$$

**2.3.2 Local Interpretable Model-agnostic Explanations (LIME)**

Local Interpretable Model-agnostic Explanations (LIME) approximates complex models locally by generating a new dataset of perturbed samples around the input of interest and observing how the model's predictions change [33-35]. It fits a simple, interpretable model (like linear regression) to these perturbed samples, capturing the local decision boundary of the complex model. This surrogate model reveals the contribution of each feature to the specific prediction, offering insights into the model's behavior in the vicinity of the input data point. Key Components of LIME: Locality: Focuses on explaining individual predictions by approximating the model locally around the instance. Interpretable Model: Uses simple models (like linear models or decision trees) that are easy to understand. Model-agnostic: Can be applied to any ML model, regardless of its complexity.

Given a complex model f and an instance $x$, LIME aims to find an interpretable model $g$ that approximates $f$ in the vicinity of $x$. The local surrogate model $g$ is typically a linear model or a decision tree, (surrogate model refers to a simpler, interpretable model that is used to approximate and explain the behavior of a more complex, often opaque, model). The goal is to minimize the following objective function [33]:

$$\xi(f, g, \pi_x) = \sum_{z \in Z}^n \pi_x(z)(f(z) - g(z))^2 + \Omega(g), \tag{14}$$

Here $Z$ is the set of perturbed samples around $x$ and $\pi_x(z)$ denote the proximity measure that defines the locality around $x$. Typically, a kernel function is used to weigh the instances based on their distance to $x$. $f(z)$ is the prediction of the complex model, $g(z)$ is the prediction of the surrogate model, and $\Omega(g)$ is a complexity term to ensure that $g$ remains interpretable.

LIME works by Creating a dataset of perturbed samples $Z$ around the instance $x$. Perturbations are often generated by sampling features around their values in $x$. Next it computes the predictions by complex model $f$ to predict the outcomes for the perturbed samples $Z$. It applies a proximity measure $\pi_x(z)$ to weigh the perturbed samples based on their distance to $x$. A common choice is the exponential kernel [34]:

$$\pi_x(z) = exp(-\frac{D(x,z)^2}{\sigma^2}), \tag{15}$$

where $D$ is a distance metric (e.g., Euclidean distance) and $\sigma$ controls the width of the neighborhood.

**2.3.3 SHapley Additive exPlanations (SHAP)**

SHapley Additive exPlanations (SHAP) calculates the contribution of each feature to a prediction using Shapley values from cooperative game theory (see figure 3), which ensures fair attribution [36],[37],[38]. This involves considering all possible combinations of feature subsets and computing the change in the prediction when each feature is added. By averaging these changes across all possible subsets, SHAP provides an exact and fair distribution of feature importance for each prediction, allowing for both local and global interpretability.

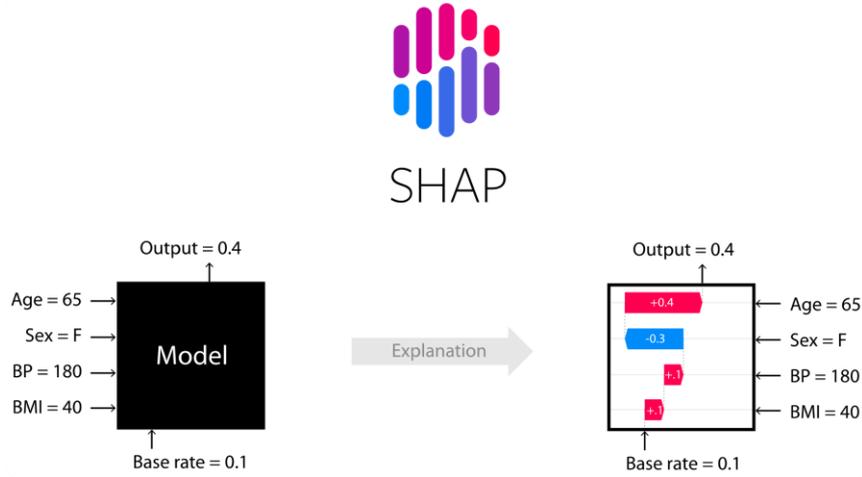

**Figure 3.** SHAP for explain ML predictions. A local explanation based on assigning a numeric measure of credit to each input feature

Value can be dined as for a model $f$ and a feature $i$, the Shapley value $\phi_i$ represents the contribution of feature $i$ to the prediction. The Shapley value is defined as [36]:

$$\phi_i(f,x) = \sum_{S \subseteq N \setminus \{i\}}^{n} \frac{|S|!(|N|-|S|-1)!}{|N|!} [f(S \cup \{i\}) - F(S)], \tag{16}$$

Where $N$ is the set of all features. S is a subset of features not including $i$. $f(S)$ is the prediction made using the subset of features S. $|S|$ is the number of features in subset $S$. The SHAP values efficiency are the sum of the SHAP values for all features equals the difference between the prediction $f(x)$ and the average prediction (baseline) [37]:

$$\sum_{i=1}^{n} \phi_i(f,x) = f(x) - f(x)', \tag{17}$$

where x′ is the baseline input (typically the average prediction over the training data).

**2.3.4 Integrated Gradients (IG)**

Integrated Gradients (IG) attributes an ANN's prediction to its input features by integrating the gradients of the output with respect to the input features along a path from a baseline input to the actual input [39]. This method accumulates the contributions of each feature along the integration path, providing a comprehensive measure of each feature's influence on the prediction. The result is a set of attribution scores that indicate the contribution of each input feature to the prediction.

For a $F$ (representing the neural network) and input $x$ with baseline $x'$, the integrated gradient for the $i - th$ feature is given by [39]:

$$IntegratedGrad_i = (x_i - x_i') \int_{\alpha=0}^{1} \frac{\partial F(x_i' + \alpha((x_i - x_i'))}{\partial x_i} d\alpha, \tag{18}$$

Here $x_i$ is the $i - th$ feature of the input, $x_i'$ is the $i - th$ feature of the baseline input, $\alpha$ is a scalar that ranges from 0 to 1, creating a path from $x'$ to $x$, and $\frac{\partial F}{\partial x_i}$ is the gradient of the model output with respect to the $i - th$ input feature.

**2.3.5 Partial Dependence Plots (PDP)**

Partial Dependence Plots (PDP) visualize the relationship between one or two features and the predicted outcome, holding other features constant [40]. By plotting the average predicted response as a function of the feature values, PDPs illustrate how changes in specific features affect the model's predictions. This method helps highlight nonlinear relationships and interactions between features, providing intuitive insights into the model's behavior and the influence of particular features.

Given a model $f(x)$ where $x = (x_1, x_2, ..., x_p)$ represents the feature vector, the partial dependence of the prediction on a particular feature $x_i$ (or a set of features $x_s$) is defined as the expected value of the model output over the distribution of the other features.

For a single feature $x_j$:
The partial dependence function $\widehat{f_{x_j}}(x_j)$ is given by:

$$\widehat{f_{x_j}}(x_j) = \frac{1}{n}\sum_{i=1}^{n} f(x_j, x_iC), \tag{19}$$

where $x_iC$ represents all the other features except $x_j$ for the $i-th$ observation.

### 2.3.6 Permutation Feature Importance (PFI)

Feature importance methods quantify the impact of each feature on the model's predictions. One common approach, permutation importance, involves shuffling a feature's values and measuring the change in model performance (as shown in figure 4). A significant drop in performance indicates high feature importance. Other methods include analyzing the weights of linear models or the depth of splits in decision trees. The output is a ranking or score for each feature, showing how influential each one is in determining the model's predictions. Permutation Feature Importance (PFI) was introduced by Breiman in 2001 for RF [41]. Based on this idea, [42] proposed a model-agnostic version of the feature importance and called it model reliance [43]. This approach, also known as the Mean Decrease Accuracy (MDA), can be used as an alternative to overcome the drawbacks of default feature importance computed with Mean Decrease in Impurity (MDI) [44].

PFI modifies the feature values and evaluates the increase in the prediction error of the model after permuting or shuffling a specific feature's values. This global, model-agnostic technique evaluates the impact on the expected outcome, positive or negative when the feature is permutated. It indicates a feature as important if the error increases by shuffling the feature's values a specific number of times. Otherwise, if the error does not change by shuffling the feature's values, the feature is "unimportant" [45]. A great change in accuracy after the shuffling indicates that the feature is important. If shuffling one column does not result in the model's accuracy, then the permutation score of the feature is lower [46], [47], [48], [49]. Furthermore, the PFI technique benefits from being model-agnostic and can be calculated many times with different permutations, allowing a straightforward interpretation of the results. However, it is computationally expensive and might have problems with highly correlated features [44].

Algorithm 1: The main steps of the PFI based on Fisher, Rudin, and Dominici (2018).

| 1 | | Estimate the original model error $e^{orig} = L(y, f(X))$ (for example using the mean squared error (MSE)) |
|---|---|---|
| 2 | | For each feature $j = 1, ..., p$ do: |
| | 2.1 | Generate feature matrix $X^{perm}$ by permuting or shuffling feature j in the data X. This breaks the association between feature j and true outcome y. |
| | 2.2 | Estimate error $E^{perm} = L(y, f(X^{perm}))$ based on the predictions of the permuted data. |
| | 2.3 | Calculate PFI using one of the following expressions: $$PFI^j = \frac{E^{perm}}{E^{orig}}$$ $$PFI^j = E^{perm} - E^{orig}$$ |
| 3 | | Sort features by descending PFI. |

The main steps of the PFI based on [42] are illustrated in Algorithm 1 where $f$ is the trained model, $X$ is the feature matrix, $p$ the number of features, $y$ is the target vector and $L(y, f)$ is the error measure [43].

### 2.3.7 Counterfactual Explanations (CE)

Counterfactual explanations show how to change the input to achieve a different prediction [50]. This method identifies the minimal changes required in the input features to reach a desired outcome. It involves searching for a new set of feature values that, when input to the model, produces the desired prediction. Optimization techniques are often used to find the closest valid counterfactual, providing an explanation of what needs to be altered in the input to change the prediction. Given a model f that maps input x to an output y, a counterfactual explanation aims to find a modified input x' such that:

$$f(x') \neq f(x), \tag{20}$$

where $x'$ is as close as possible to $x$ according to some distance metric $d(x, x')$.

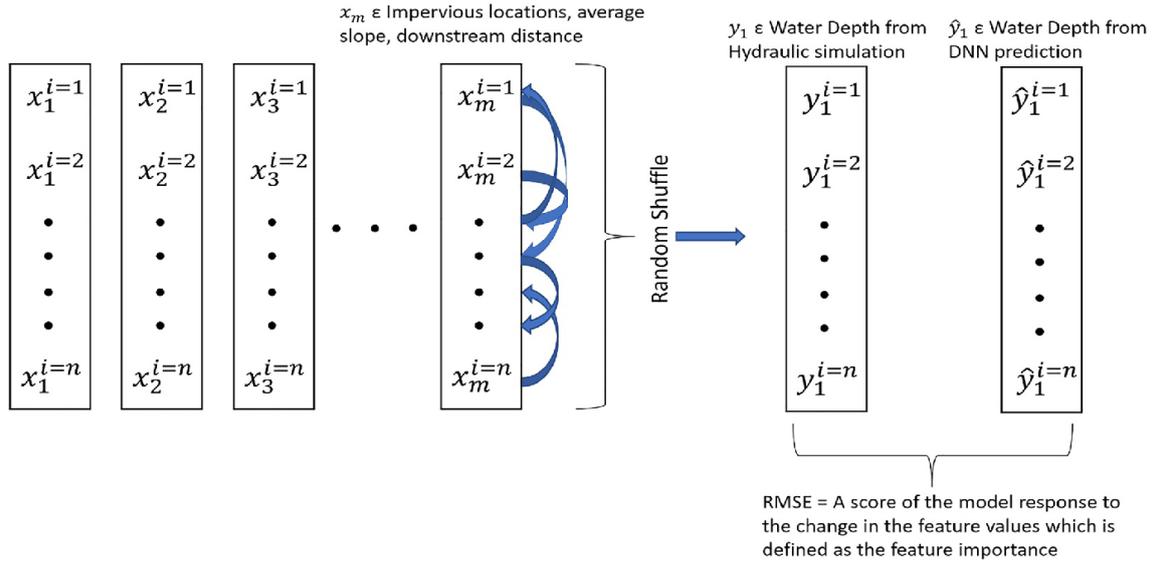

**Figure 4.** Mechanism of PFI, as illustrated the feature's values are shuffled and measured based on the change in model performance.

**2.3.8 Saliency Maps (SM)**

Saliency maps (SM) play a key role in the field of XAI by identifying the way AI models make decisions [51]. These maps focus on the salient features of an image, helping in visual cognition and attention. They highlight the input features of the algorithm in its decision-making process, aiding in the interpretation of complex AI models. Incorporating human attention knowledge into saliency-based XAI methods enhances the plausibility and faithfulness of explanations. Saliency maps are a visualization technique used in ML, particularly in CNNs, to highlight the regions in an input image that are most relevant to the model's prediction. The concept relies heavily on the gradients of the model's output with respect to the input.

Given a model $f(x)$, where $x$ is the input image and $f(x)$ is the output (such as a class score in classification), the saliency map $S(x)$ is computed as:

$$S(x) = \frac{\partial f(x)}{\partial x}, \tag{21}$$

Here The partial derivative derivative $\frac{\partial f(x)}{\partial x}$ measures how much the output $f(x)$ changes with respect to a small change in the input $x$. This tells us which pixels in the input image are most influential for the prediction.

**2.3.9 Gradient-weighted Class Activation Mapping (Grad-CAM)**

Grad-CAM provides visual explanations for CNNs by computing the gradients of the output class score with respect to the feature maps of the last convolutional layer. These gradients highlight the importance of each neuron for class prediction [52]. By averaging the gradients to obtain weights and creating a weighted sum of the feature maps, Grad-CAM produces a heatmap that shows which regions of the input image are most influential for the prediction, offering a visual interpretation of the model's focus.

For CNN, let $A^k$ be the $k-th$ feature map of the last convolutional layer, and $y^c$ be the score for class $c$. The Grad-CAM for class $c$ c is computed as follows:

$$Compute\ Gradients = \frac{\partial y^c}{\partial A^k}, \tag{22}$$

$$Global\ Average\ Pooling = \alpha_k^c = \frac{1}{z}\sum_i^n \sum_j^n \frac{\partial y^c}{\partial A^k}, \tag{23}$$

Here, $z$ is the number of pixels in the feature map $A^k$, and $\alpha_k^c$ is the importance weight for the $k-th$ feature map.

$$Weighted\ Sum: L_{Grad-CAM}^c = ReLU\left(\sum_k^n \alpha_k^c A^k\right), \tag{24}$$

Here, $L_{Grad-CAM}^c$ is the Grad-CAM heatmap for class $c$.

**2.3.10 Intrinsic and post-hoc methods Advantages and disadvantages**

In the field of Explainable Artificial Intelligence (XAI), methods are categorized into intrinsic and post-hoc approaches, each with advantages and disadvantages. Intrinsic XAI methods are interpretable by design. For example, Partial Dependence Plots (PDPs) show relationships between features and outcomes [240], and Permutation Feature Importance (PFI) evaluates feature importance by shuffling input values [240]. These methods are simple and easy to interpret but struggle with complex, non-linear patterns [238].

Post-hoc XAI methods, applied after model training, explain complex models. Layer-wise Relevance Propagation (LRP) traces model predictions back through layers [239], while LIME (Local Interpretable Model-agnostic Explanations) approximates local model behavior [4]. SHAP (SHapley Additive exPlanations) assigns feature importance using game theory [240], and Integrated Gradients (IG) attributes contributions through gradient integration [239]. Visual tools like Saliency Maps (SM) and Grad-CAM highlight important regions in input data [238]. Counterfactual Explanations (CE) identify minimal input changes that alter predictions [240]. Post-hoc methods are flexible and suitable for any model but can be computationally expensive and provide approximations [238][240].

Post-hoc interpretability methods like SHAP (Shapley Additive Explanations) and LIME (Local Interpretable Model-agnostic Explanations) provide valuable insights into black-box models, but they suffer from challenges related to statistical stability. Both methods rely on approximations and random sampling, which can lead to variability in the results, especially when the underlying model is unstable or sensitive to data changes. LIME fits a local surrogate model based on data perturbations, but small changes in the data can lead to different explanations for the same prediction, introducing instability. Similarly, SHAP calculates feature contributions based on Shapley values, but the approximations used in large datasets or high-dimensional models can cause inconsistencies. Both methods can also struggle with noise and outliers in the data, and their performance can degrade when features interact non-linearly. These challenges highlight the need for caution when interpreting the results of post-hoc methods, with additional stability checks recommended [242][243].

Integrating Explainable AI (XAI) with AI/ML algorithms in clinical workflows holds great potential to enhance trust, transparency, and decision-making in healthcare. XAI methods, such as SHAP and LIME, can explain complex model predictions by highlighting key features or patient data points that contributed to a diagnosis or treatment recommendation, helping clinicians understand and validate AI outputs. To effectively integrate XAI into clinical settings, it is crucial to design models that balance transparency with predictive performance, ensuring real-time, patient-centered explanations that align with clinicians' decision-making processes. XAI should be incorporated into clinical decision support systems (CDSS), where it can provide interpretable, actionable insights in real-time, supporting clinicians without overwhelming them with excessive information. However, challenges such as data quality, model instability, regulatory compliance, and clinician adoption must be addressed. XAI must be validated through rigorous clinical trials to ensure its relevance, accuracy, and reliability in diverse patient populations. By fostering interdisciplinary collaboration between AI experts, clinicians, and ethicists, and by ensuring regulatory adherence (e.g., HIPAA, GDPR), XAI can be integrated seamlessly into clinical workflows, enhancing both the quality of care and patient outcomes. Case studies in radiology, predictive analytics, and clinical decision-making already demonstrate how XAI can improve trust and efficacy in AI-driven healthcare, though challenges like data fragmentation and clinician resistance to new technologies remain. Ultimately, XAI can bridge the gap between AI's powerful capabilities and its responsible use in healthcare.

**3. Methods**

The methodology of this review paper is designed to provide a comprehensive examination of the existing literature on the application of XAI in human health monitoring as illustrated in Figure 5. Our approach involves several key steps to ensure a thorough and balanced overview of the topic in the following subsections.

*3. 1 Literature Search Strategy*

The primary source of literature for this review was a comprehensive search of electronic databases including PubMed, IEEE Xplore, Web of science, Google scholar, Nature, Research Gate, Science Direct and Scopus as shown in Figure 5. Initial search on XAI in human health monitoring reviled several diseases that have been trending on most electronic databases therefore a list of keywords used in the search included:

- XAI Human health
- XAI Gait

- XAI Parkinson.
- XAI stroke.
- XAI depression.
- XAI cancer.
- XAI Heart Disease.
- XAI Alzheimer's Disease.
- XAI COVID-19

The search was limited to articles published in English between 2014 and 2024 to focus on the most recent advancements in the field. The initial search yielded 478 items, after careful evaluation of the abstract, review papers and unrelated papers were excluded resulting in 194 references. New keywords were identified based on the XAI algorithms as:
1. LRP (Layer-wise Relevance Propagation)
2. LIME (Local Interpretable Model-agnostic Explanations)
3. SHAP (SHapley Additive exPlanations)
4. IG (Integrated Gradients)
5. PDP (Partial Dependence Plot)
6. PFI (Permutation Feature Importance)
7. CE (Counterfactual Explanations)
8. SM (Saliency Maps)
9. Grad-CAM (Gradient-weighted Class Activation Mapping)

Each of the above keywords was searched in combination with one of the following terms: Gait, Parkinson, stroke, depression, cancer, Heart Disease, Alzheimer's Disease, or COVID-19. This search process resulted in a total of 259 research papers. Figure 5 illustrates the different stages of the search process, ranging from keyword refinement to the application of inclusion and exclusion criteria. The inclusion criteria for articles were narrowly defined to focus on discussions of XAI in health monitoring and studies that provided concrete, empirical data on its real-world effectiveness. Articles not directly related to XAI or health monitoring, such as those that emphasized medical devices, review papers, or theoretical concepts, were excluded from the analysis to ensure relevance and focus.

### 3.2 Selection and Data Extraction Process

The selection process, as outlined in Figure 5, adhered strictly to a set of predefined inclusion and exclusion criteria to ensure that the selected studies were relevant to the application of XAI in human health monitoring. Articles were included if they specifically discussed XAI applications in healthcare settings and provided empirical data on the effectiveness of XAI tools. Studies were excluded if they focused solely on non-XAI-related technologies, such as medical devices, or if they were review articles lacking empirical data.

Following the selection, relevant data were systematically extracted from each article, including the XAI technique employed, the specific health condition it addressed, its demonstrated or potential impact on health monitoring, and any challenges or limitations discussed. The extracted information was then categorized into thematic areas based on the XAI methodologies, such as LIME, SHAP, and Grad-CAM, and their application in various medical conditions like stroke, Parkinson's disease, and cancer.

### 3.3 Quality Assessment and Synthesis of Information

To ensure the reliability and credibility of the selected sources, a rigorous quality assessment was carried out. This included evaluating the reputation of the publication venue, the credibility of the authors, and the methodological soundness of the studies. Only high-quality studies that met these criteria were included in the final literature pool. The synthesized data from the selected studies were analyzed to provide a com-prehensive overview of the current landscape of XAI in health monitoring. Common themes and trends were identified, along with gaps in the existing research that point to future areas of investigation. The final synthesis aimed to highlight the most impactful applications of XAI in healthcare and to propose directions for future research and development in the field.

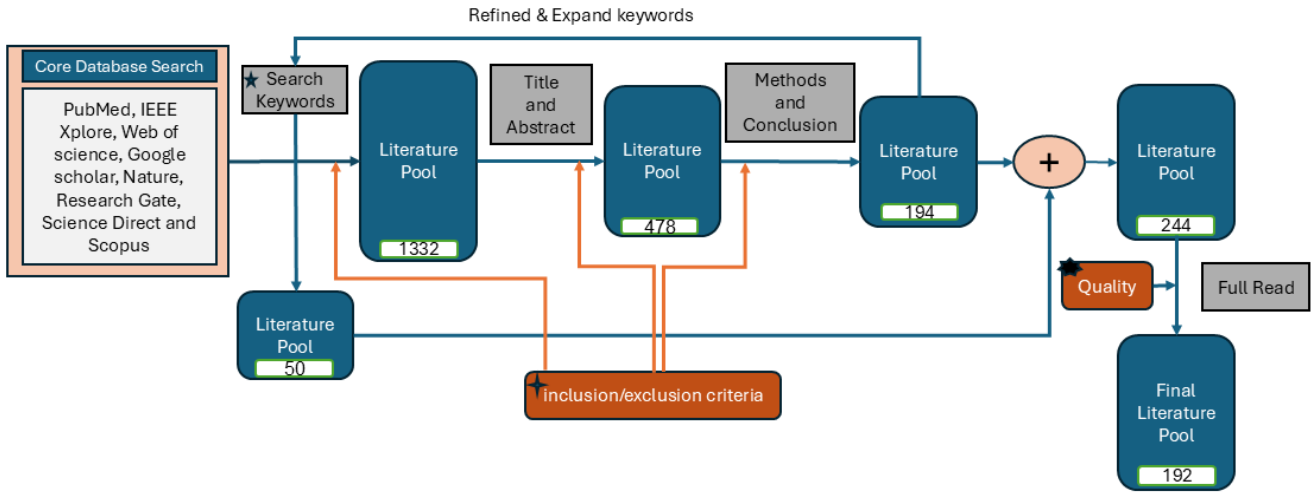

**Figure 5.** The systematic literature review process. Search keywords and inclusion/exclusion are listed in article selection section.

## 4. Review Findings

This section presents the findings from the literature review focused on XAI techniques used in the diagnosis of eight specific diseases. These diseases were selected due to their prevalent reliance on ML and XAI for diagnosis and monitoring. The diseases include Human Gait, Parkinson's Disease, Depression, Cancer, Heart Disease, Alzheimer's Disease, and COVID-19. The subsequent sections will organize the literature findings based on the widely used XAI methods for diagnosing these diseases.

### 4.1. Layer-wise Relevance Propagation (LRP)

LRP has been widely integrated into human health monitoring, as evidenced by our comprehensive literature review. This method is applied across various medical domains, including gait analysis [53-58], cancer [59-65], COVID-19 [66-70] cardiovascular health [71-74], Parkinson's disease [75-77], Alzheimer's disease [78-80], and mental health [81-86].

In the context of human gait analysis, several studies have employed LRP to enhance the interpretability of DL models. For instance, LRP was used in [53] to understand gait patterns through floor sensor data, where Deep Neural Networks (DNNs) were applied to predict cognitive load during dual-task gait scenarios and identify key features contributing to the classification. Similarly, in [54], researchers utilized ground reaction forces and full-body joint angles with DNNs to elucidate individual gait patterns, focusing on characterization and identification, while LRP facilitated the interpretation of model predictions, as depicted in Figure 6. Additionally, in [55], LRP was implemented to explore the detection and severity rating of freezing of gait (FOG) episodes in Parkinson's disease patients using CNNs and kinematic data, highlighting reduced peak knee flexion and fixed ankle dorsiflexion during the swing phase as significant features.

Further applications of LRP include its use in [56] with CNNs and wearable inertial sensor data to identify gait modifications under fatigue and cognitive task performance, where LRP visualizations revealed highly individualized gait changes and precise time-step alterations in the input signal. In [57], LRP was applied to investigate gait pattern differences between high- and low-mileage runners, suggesting that ankle and knee data provide substantial information for recognizing gait features, particularly in the sagittal and transverse planes. Moreover, [58] examined smartphone-based sensors to characterize ambulation in Multiple Sclerosis (MS), with interpretable DL methods like LRP offering insights into the role of cadence-based measures, gait speed, and ambulation-related signal perturbations in distinguishing MS disability from healthy participants. Finally, [59] focused on analyzing spatiotemporal gait signals for Parkinson's disease detection and severity rating, using LRP to interpret model outputs and identify the most significant features in the spatiotemporal ground reaction force (GRF) signals.

In cancer research, LRP has been consistently used to enhance the interpretability of ML models across various applications. For instance, [60] explores morphological and molecular profiling of breast cancer using explainable ML techniques on histology images, with the goal of improving diagnostic accuracy and informing treatment planning. Similarly, [61] focuses on predicting estrogen receptor status from hematoxylin-eosin images using

DNNs, contributing to more personalized breast cancer therapy strategies. Additionally, [62] addresses the challenges in analyzing histopathological images for tumor tissue discrimination by employing pixel-wise heatmap visualizations and bias reduction methods to improve interpretability.

Further extending the use of explainable AI in cancer research, [63] investigates patient-level proteomic network predictions, using explainable AI to infer protein interaction networks and cancer-type specificity. In a related application, [64] applies explainable AI to predict gene regulatory networks from single-cell RNA sequencing data, providing valuable insights into gene interactions relevant to lung cancer treatment. Shifting focus to hematological diseases, [65] demonstrates the feasibility of using DL for leukemia biomarker identification by employing a sequential model for malignant leukocyte detection from peripheral blood images.

During the COVID-19 pandemic, various studies have employed explainable AI techniques to improve the interpretability and accuracy of models diagnosing respiratory conditions. For instance, [66] used explainable augmented intelligence and deep transfer learning for pediatric pulmonary health evaluation from chest X-ray images, demonstrating how AI can aid in diagnosing respiratory conditions in children. Similarly, [67] developed Deep-COVID-Explainer, a model for COVID-19 diagnosis from chest X-ray images that emphasizes visual explanations to enhance diagnostic accuracy. In another study, [68] introduced an explainable AI model for lung disease classification using LRP to interpret decisions based on chest X-ray images. Additionally, [69] proposed COPDNet, an explainable ResNet50 model designed for diagnosing chronic obstructive pulmonary disease (COPD) from chest X-rays, which enhances model trustworthiness and interpretability in clinical settings. Further, [70] presented M-Rule, an enhanced deep Taylor decomposition method for multi-model interpretability in COVID-19 classification from lung CT images, providing detailed relevance propagation insights for accurate disease classification.

Shifting focus to cardiovascular health, [71] introduced CureNet, a model that leverages custom large language models to interpret ECG signals and diagnose cardiac abnormalities, thereby supporting comprehensive clinical decision-making. Meanwhile, [72] explored stroke outcome prediction using Multilayer Perceptrons (MLPs) and tree boosting models, focusing on feature importance and interpretability to enhance clinical outcome predictions. Another study, [73], introduced xECGArch, a DL architecture that offers trustworthy and interpretable ECG analysis, considering both short-term and long-term features to diagnose cardiovascular diseases. Lastly, [74] analyzed a DL model for 12-Lead ECG classification, revealing learned features akin to diagnostic criteria for conditions like atrial fibrillation and left bundle branch block, thereby enhancing ECG pattern interpretation in clinical practice.

In the context of neurodegenerative diseases, [75] applied RNNs to identify frequency characteristics of tremor types in Parkinson's disease using explainable AI, which aids in better understanding and managing motor symptoms. Additionally, [76] utilized CNNs for the automatic classification of dopamine transporter SPECT images, improving both diagnostic accuracy and interpretability in distinguishing Parkinsonian syndromes. Another study, [77], identified diagnostically useful extra-striatal signals in dopamine transporter SPECT using 3-dimensional CNNs and explainable AI, emphasizing the importance of visual explanations beyond traditional striatal signals in Parkinson's disease diagnosis.

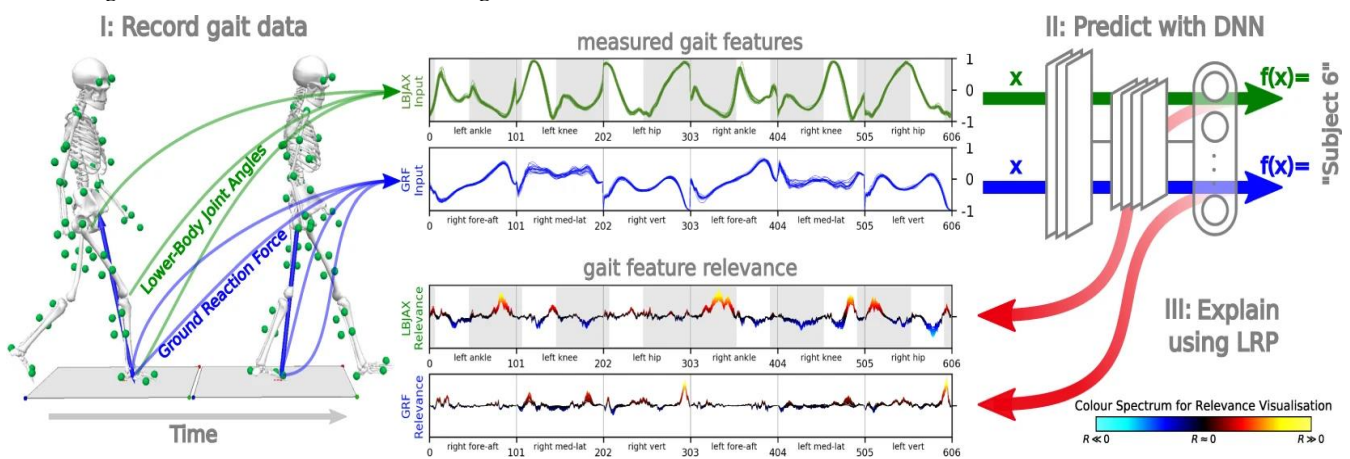

**Figure 6.** LRP for gait relevance using CNN and LRP to explain the identification of subjects based on their gait captured with force plates, and ten infrared cameras recorded the three-dimensional full body joint angles and ground reaction forces (GRF) [54].

Regarding Alzheimer's disease, [78] proposed a region-based CNN framework for Alzheimer's treatment planning, integrating tabular data, MRI images, and genetic information to classify disease stages and plan personalized therapies. Similarly, [79] used CNNs to capture patterns of Alzheimer's disease from T1 MRI scans, employing heatmap visualizations to elucidate structural changes associated with disease progression. In another study, [80] investigated structural changes in the aging brain using CNNs for brain-age estimation, contributing to early detection of neurodegenerative disorders and personalized treatment planning. Furthermore, [81] compared heatmap methods with meta-analysis maps to capture Alzheimer's disease effects on the brain, using SVM, CNNs, and ResNets to improve diagnostic accuracy and interpretability.

In the field of mental health, [82] explored brain texture as a marker for transdiagnostic clinical profiles in psychosis and depression, using explainable AI algorithms to link brain texture changes with psychopathological states. Additionally, [83] applied DNNs for interpretable single-trial EEG classification, enhancing the understanding of neural correlates in mental health conditions and cognitive processes. [84] proposed a novel brain connectivity score derived from fMRI data using explainable neural networks, providing insights into brain functional integration. Moreover, [85] integrated LRP with medical imaging to enhance diagnostic accuracy and interpretability, showcasing AI's transformative impact on healthcare. Finally, [86] proposed advanced AI models for medical image analysis, emphasizing the role of interpretable ML in improving clinical outcomes. The utilization of DNNs to explore brain connectivity and identify biomarkers in neurological and psychiatric conditions.

*4.2. Local Interpretable Model-agnostic Explanations (LIME)*

This section highlights the use of LIME in human health monitoring, showcasing innovative applications, addressing notable challenges, and exploring promising future directions across various domains, including gait analysis [87-91], cancer [92-97], COVID-19 [68],[98-104], Parkinson's disease [105-108], Alzheimer's disease [78],[109-113], mental health [114-116], and stroke [117-121].

The application of LIME has significantly improved the transparency and interpretability of ML models in gait analysis, Özateş et al [87]. manually developed novel gait analysis features and employed AI models such as SVM and RF to remove low-variance features. They then used SVM, RF, KNN, and LR for classification, with LIME identifying features influencing gait patterns and foot conditions, demonstrating how LIME elucidates model decisions. Similarly, Dutta and Puthal [88] applied LIME and SHAP at Cloud and Edge levels to clarify AI model predictions for cardiac arrest, enhancing health status awareness. In another study [89], LIME and SHAP improved the accountability of classification models (e.g., Decision Trees, Gaussian Naive Bayes, RF, LGBM, and KNN) in predicting fetal health outcomes. SHAP assessed feature contributions while LIME analyzed specific instances, enhancing interpretability. Furthermore, in [90], an interpretable ML framework using Electromyography (EMG) data distinguished between stroke patients and healthy individuals, employing SHAP, LIME, and Anchors to identify significant EMG spectral features related to specific muscles. Additionally, [91] introduced a ML approach using a RF model to predict upper limb recovery scores post-stroke rehabilitation, with Explainable AI techniques elucidating influential prediction variables, aiding clinical management of sub-acute stroke patients.

LIME has consistently advanced cancer research by making AI model predictions more interpretable. For instance, [92], introduced an explainable deep neural network model, NeuroNet19, for brain tumor classification, using LIME to highlight key predictive features. Duell et al. [93] conducted a comparative analysis of three explainable AI methods, including LIME, to elucidate mortality predictions in lung cancer patients using Electronic Health Records (EHRs), emphasizing feature importance in clinical decisions. Hasan et al. [94] employed LIME to interpret DL models like ResNet50 in classifying brain tumors from MRI images, identifying crucial image regions and enhancing decision transparency. Another study used LIME to analyze breast cancer metastasis, employing a CatBoost classifier to address classification bias [95]. Additionally, integrated LIME and SHAP to understand ML models' decisions in breast cancer diagnosis, identifying key predictive factors [96]. Research on breast cancer recurrence combined SHAP and LIME with models such as SVM, KNN, NB, LG, and XGB, highlighting influential recurrence risk factors and aiding personalized treatment plans [97].

During the COVID-19 pandemic, LIME and other XAI techniques enhanced model interpretability in medical diagnostics. Pitroda et al. [68] developed a framework using customized CNNs for lung disease identification from chest X-rays, integrating LIME, LRP, and Deep Taylor Decomposition (DTD). Another notable study utilized explainable ML techniques to predict COVID-19 using various imaging modalities, emphasizing model transparency [98]. The innovative CoughLIME generated signified explanations from COVID-19 cough data, offering audio-based interpretability [99]. The NasNetMobile model, incorporating LIME, provided visual explanations to enhance COVID-19 symptom detection decisions [100]. Another study employed LIME and Guided Backpropagation (GB) for lung disease classification from chest X-rays, highlighting model transparency [101]. Explicable AI techniques also aided COVID-19 surveillance and interpretation using X-ray imaging [102].

Another study evaluated XAI methods like GradCAM, LIME, RISE, and gradient-based approaches to detect biases in COVID-19 CT-scan classifiers [103]. Lastly, the LISA framework unified current XAI techniques with LIME to enhance medical image explainability [104].

Recently, LIME has become essential in medical diagnostics, particularly for Parkinson's disease prediction. Nguyen and Byeon [105] developed a LIME-based stacking ensemble model to predict depression in Parkinson's disease patients, integrating Logistic Regression with base models (LightGBM, KNN, RF, Extra Trees, AdaBoost), enhancing prediction accuracy and interpretability. Magesh et al. [106] trained a CNN model on DaTSCAN imagery to classify Parkinson's patients, using LIME to provide visual indicators of important regions, crucial for clinical settings. Saravanan et al. [107] developed an EXAI model using LIME to analyze spiral and wave drawings for early Parkinson's detection, enhancing model accuracy and providing diagnostic insights. The authors of [108] proposed a ML pipeline using time series data and high-performance models (SVM, RF, LR, DT, ET, GNB, LGBM, SGD, ADB, KNN), leveraging XAI models LIME and SHAPASH for decision insights.

Transitioning to Alzheimer's disease (AD), Parvin et al. [78] used SVM to classify AD patients, with LIME interpreting prediction insights and identifying significant demographic features. Another study [109] utilized EEG signals to predict AD development, with LIME explaining CNN model diagnoses. A separate investigation employed classifiers like SVC, KNN, and XGBoost with gene expression data for AD classification, with LIME elucidating significant genes [110]. Researchers also analyzed MRI data using CNNs to identify initial AD patterns, with LIME revealing visual evidence supporting model predictions [111]. Another study proposed a Bayesian-optimized hybrid DL model with explainability features for early AD detection, using LIME for individual prediction explanations [112]. Lastly, [113] used Neural Network models to detect early-stage AD, with LIME pinpointing specific regions associated with classifications.

In mental health, LIME has elucidated depression predictions from various ML models (e.g., LR, NB, kNN, SVM, DT, RF, AdaBoost, XGBoost), enhancing clinical understanding [114]. An ensemble model combining SVM and kNN addressed depression and suicidal ideation classification, using LIME to identify influential classification keywords as shown in Figure 7 [115]. A novel approach classified depression from speech using a CNN and LSTM model, with LIME pinpointing significant spectrogram parts [116].

For stroke prediction, LIME interpreted multiple ML models (LR, SVM, RF, XGB, ANN), providing comprehensive prediction insights [117]. Mridha et al. proposed an automated stroke prediction algorithm, using SHAP and LIME to understand model decisions [118]. Another study used LIME to explain Adaptive Gradient Boosting, XGBoost, and LightGBM models, identifying significant stroke prediction features [119]. An additional approach combined LIME with other XAI models for stroke prediction explanations [38]. Various classifiers (NN, SVM, RF, DT, GB) were used to classify patients with brain stroke, with LIME being utilized to interpret the best-performing model, the RF [120]. Lastly, the application of LIME across different models consistently highlighted the substantial impact of specific features on predicting the occurrence of stroke [121].

*4.3. Shapley Additive Explanations (SHAP)*

This This section focuses on the integration of SHAP in human health monitoring. SHAP is introduced in [1], and its values are utilized to measure the output of the ML model [1]. It selects the individual predictions and the important value for each feature for a certain prediction [122]. SHAP was employed in many studies related to human disease diagnosis, including gait and geriatric analysis [122], pneumonia infection [123], cancer [124-136], Alzheimer's disease [79],[112], [137-150], stroke disease [38],[118],[72],[90], [151-153], Parkinson's Disease [108],[154-157], depression [158 - 160], heart disease [161-163], and Covid-19 [164-168].

Related to human gait and geriatric analysis, SHAP was used to identify the best model to predict the fall risk for older adults [122]. It was shown that the SVM model with a linear kernel was the best predictor for 6-month fall risk. In [123], XAI was used to detect pneumonia infection and provide feedback about the quality predictions. Also, SHAP provides the visual representation of the low and high pneumonia infections on various Chest X-ray (CXR) datasets.

XAI to detect and predict prostate cancer was studied in [124] and with the use of SHAP it was found that the RF model has higher performance with more accuracy compared to other approaches across all datasets. In [125], SHAP also was used to understand Raman spectra for thyroid cancer diagnosis, employing several algorithms to help predict the thyroid lesion's malignancy. Similarly, breast cancer was investigated in [126] with the use of XAI and ML resulting in higher prediction for cancer diagnosis. Additionally, SHAP identified the LightGBM model has higher performance and accuracy compared to other approaches. In [127], XAI and SHAP were employed to predict colorectal cancer using gut microbiota data and demographic information. It was proven that RF outperforms XGBoost and SVM models for predicting colorectal cancer. In [128], Several ML models using Shaply XAI were studied to improve ovarian cancer predictions. It was shown that the proposed stacking classifier system combined various ML techniques to attain superior prediction performance through SHAP analysis. XAI models containing SVM-SHAP, XGBoost, and RF were used in [129] to predict ovarian cancer, and SHAP was employed to select the best 30 feature genes. Research within XAI and ML models, particularly the SHAP framework, was studied in [130]-[136] to identify cancer disease. Those studies had the same objective but varied in detection methods, dataset balancing, and feature selection.

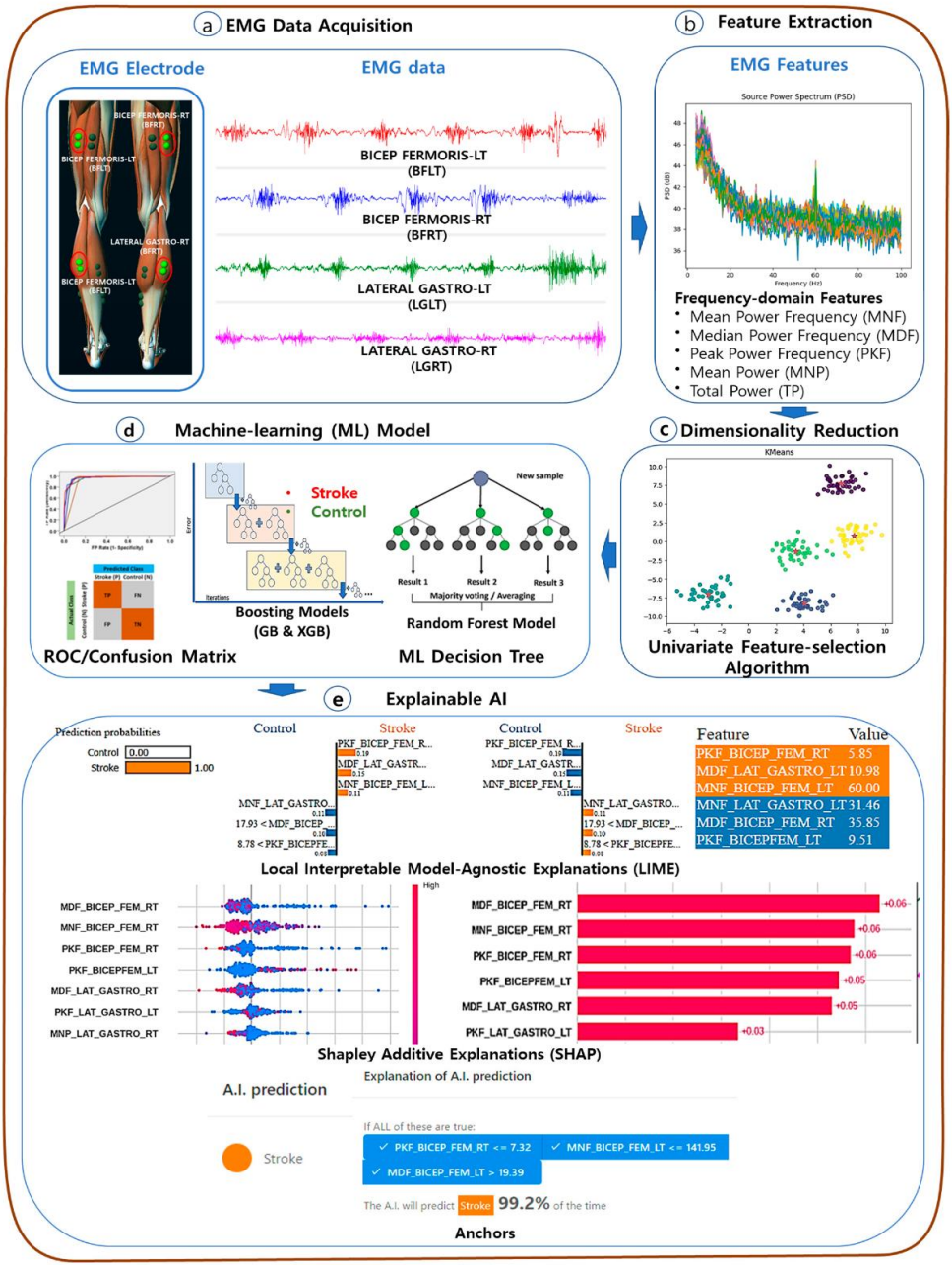

**Figure 7.** Diagram illustrating an explainable EMG-based model for predicting stroke-impaired gait. It includes steps for EMG data acquisition, feature extraction, dimensionality reduction, machine learning model evaluation, and the application of XAI methods (LIME, SHAP, Anchors) to interpret model predictions [90].

In Alzheimer's disease, a highly accurate XAI with RF model was proposed in [137] to detect and predict Alzheimer's disease. The proposed RF model based on SHAP has better prediction performance compared to other models. Similarly, XAI based on RF was employed in [138] to study the role of cognitive and clinical measures in the automated classification of mild cognitive impairment and Alzheimer's disease. SHAP values here are used to measure the impact of each index accurately to follow the cognitive status over time and obtain information about Alzheimer's disease. In [139], XAI with XGBoost proposed model outperforms the RF model and predicts Alzheimer's disease with higher accuracy. The proposed model was explained by SHAP values to diagnose Alzheimer's disease in the early stage. Additionally, the effect of brain connectivity due to Alzheimer's disease was examined in [140], and it was shown that SHAP provides high robustness in employing each feature on predictions. In [141], a non-linear neural network with SHAP were investigated to detect Alzheimer's disease. The findings highlight key factors such as CDR, Age, and ASF as crucial in predicting dementia, with SHAP values offering a detailed insight into the model's decision-making process. Numerous studies from [22] to [150] and [79],[112] have explored the use of SHAP with various ML models for detecting Alzheimer's disease. These studies aim to improve the interpretability of model predictions, providing insights into the critical features influencing the diagnosis of Alzheimer's. Each research effort contributes to the growing body of literature demonstrating the effectiveness of SHAP in enhancing the transparency and trustworthiness of ML applications in medical diagnostics.

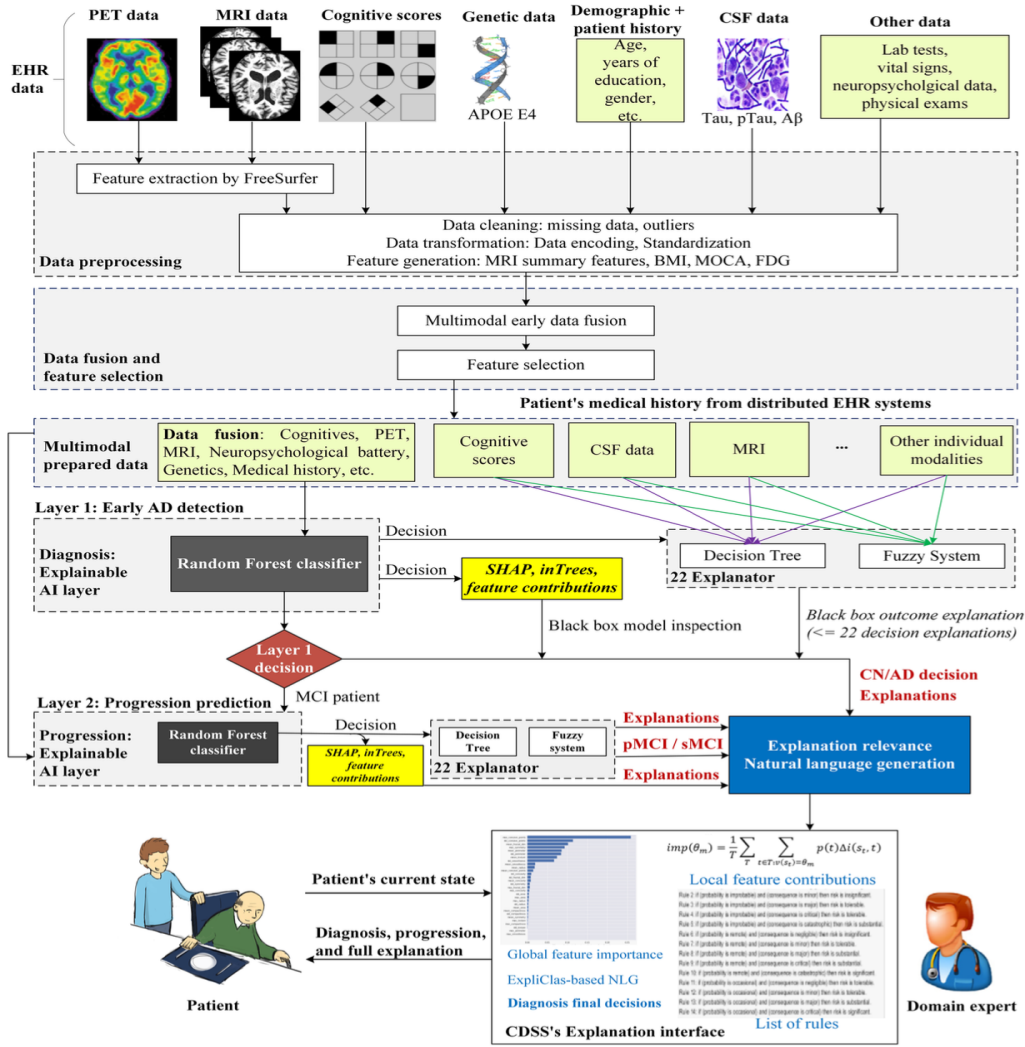

**Figure 8.** An application of SHAP framework. A variety of data modalities are used to build the predictive model. In addition, a variety of explanations are built for the entire RF behavior and for each prediction [137].

In the context of stroke analysis in [151], several XAI models were used to develop reliable predictions for stroke disease. Also, the SHAP model identified that age, body mass index (BMI), and average glucose were the main predictors for stroke in a patient. In [38], various XAI approaches were investigated to predict stroke disease using different AI models. With the use of SHAP, age, and BMI were found the most attributes to contribute to stroke disease. Similarly, in [152], SHAP based on several XGBoost models were employed for early detection of stroke disease, and it found that age, BMI, smoking state, and Average Glucose were the main features for causing stroke. Studies in [118],[72],[90],[153] have investigated stroke predictions using SHAP across various ML models. These research efforts highlight differences in detection methods, dataset balancing, and feature selection, contributing to a diverse understanding of stroke prediction and model interpretability.

Related to Parkinson's Disease a smartphone-based PD detection system was introduced in [154] that leverages interpretable prediction models, utilizing SHAP to analyze data collected from touchscreen and inertial sensors. The system evaluates features related to finger dexterity, tremor, stiffness, and hand movement, achieving over 94.5% accuracy. In [155], SHAP was applied to a model using voice and speech signal data for PD identification. This model, incorporating various preprocessing techniques and ML classifiers, achieved a maximum accuracy of 98.10%, outperforming previous studies on the same dataset. In [108], an ML pipeline for predicting Parkinson's Disease progression employed SHAP to provide explainability across five time series modalities from the PPMI dataset. LightGBM and RF models, achieving accuracies of 94.89% and 93.73%, respectively, benefited from SHAP and local explainers, which validated their medical relevance and emphasized bradykinesia as a significant feature. The study in [156] explored integrating gene expression data for PD diagnosis using SHAP to interpret the key gene features identified by logistic regression and SVM models. SHAP's insights highlighted significant biomarkers and demonstrated how XAI can support early therapeutic decisions for PD. In [157], SHAP was utilized to address high feature dimensionality in PD medical data, combined with classifiers such as gcForest, XGBoost, LightGBM, and RF. SHAP-based feature selection methods notably outperformed traditional techniques, with SHAP-gcForest and SHAP-LightGBM achieving high classification accuracy and F1-scores. This approach improved computational efficiency and offered valuable insights for diagnosing and preventing Parkinson's disease.

Consistently SHAP in depression research, [158] introduces an AI-based solution for identifying depression and explores psychiatric knowledge. It involved collecting and annotating a new audio dataset and assessing eight regression methods, with k-nearest neighbor and RF performing the best. Using SHAP, the study identified the most important features as the fourth Mel-frequency cepstral coefficients, harmonic difference, and shimmer. In [159], the study investigates ML on electronic health records (EHR) data to predict Postpartum depression (PPD) risk, enhancing screening accuracy and early detection. SHAP with a gradient tree boosting algorithm was used to identify the output of ML model. The proposed model with EPDS scores improved prediction accuracy, increasing the AUC from 0.805 to 0.844 and sensitivity from 0.72 to 0.76. In [160], SHAP was applied to early prediction of generalized anxiety disorder (GAD) and major depressive disorder (MDD) from electronic health records (EHRs). The model, trained on 59 features, achieved an AUC of 0.73 for GAD and 0.67 for MDD, with key predictors identified using SHAP values.

Moving to heart disease, [161] develops a hybrid DL model combining CNN and short-term memory (LSTM) to detect cardiovascular disease (CVD) from clinical data, achieving high accuracy rates of 73.52% and 74.15%. The model outperforms existing methods and uses explainable AI, like SHAP, to identify key CVD features, enhancing early diagnosis and screening tools. In [162], a proposed model that combines CNN with XAI to analyze heart sounds for diagnosing multiple heart conditions and assessing their severity. The model achieves a multi-label classification accuracy of 77.8% and 98% for single-label cases, using SHAP to provide insights into how diagnoses are made. This method enhances diagnostic accuracy and offers visual explanations, aiding physicians in comprehensive heart condition assessment. In [163], a DL model for detecting cardiac disorders from ECG data was introduced. The model achieved 95.8% accuracy and a 99.46% AUC, showing strong performance. The use of SHAP made the model's decisions interpretable by highlighting important changes in ECG waves, improving diagnostic clarity.

Lastly, in the Covid-19 pandemic, [164] investigate the use of XAI to examine socioeconomic factors affecting COVID-19 patient mortality. An XGBoost model was developed to forecast mortality rates, and SHAP was employed to compare global and local feature importance, revealing that Medicare financial class, age, and gender are crucial predictors. The alignment of SHAP and LIME findings highlights the effectiveness of XAI in validating feature significance. In [165], several ML models such as logistic regression, decision tree, pyFUME, and RIPPER were examined for classifying suspected COVID-19 patients using blood sample data. SHAP was used to evaluate model performance and provide interpretability through its explanations, alongside inherent features of the models. In [166], an AI with mass spectrometry (MS) data was studied to diagnose COVID-19 and employs explainable XAI to clarify decision-making. The RF classifier, achieving 94.12% accuracy, was interpreted using SHAP and other methods to create a three-stage XAI framework, helping medical practitioners understand and trust the AI model's decisions. Studies in [167] and [168] have explored the use of SHAP with various ML models for detecting COVID-19. This framework builds trust by clarifying the rationales behind the AI's decisions.

*4.4. Integrated Gradients (IG)*

Current times have seen great strides in making AI models interpretable, propping from recent achievements of XAI methods along different medical subspecialties. Parra-Mora & da Silva Cruz [169] evaluated the utility of IG in classifying epiretinal membranes alongside six other explanation techniques. This result shows the transparency of IG in providing diagnostic features for ophthalmology diagnostics, and their findings demonstrated it is able to detect interpretable insights.

Khambampati et al. [71], IG was proposed to improve the explainability of AI diagnosis with large language models. This suggests that IG could help to improve the diagnostic transparency of complex medical scenarios and make an effective management decision for patients. One example of an integrated gradient application in a hospital setting is shown in Figure 9.

Wang et al. [81] used deep neural network heatmaps, including IG, to extract patterns of Alzheimer's disease from neuroimaging investigations. Their technique, which uses interpretable AI-driven neuroimaging data, helps to better understand illness progression pathways. Singh et al. [170] created a sparse deep neural network for decoding the structural connectome, using IG to identify brain connection patterns related to neurological diseases, improving diagnosis accuracy in neurology.

Gunter et al. [171] introduced SViT (Spectral Vision Transformer) for diagnosing REM Sleep Behavior Disorder, which incorporates IG to reveal spectral properties that influence disorder categorization in sleep medicine. Barros et al. [172] investigated the interpretability of a model for automatic handwritten signature verification using IG, hence improving forensic and authentication applications through transparent AI decision-making. Nguyen et al. [173] examined longitudinal social media data to understand suicidal behavior, using IG to evaluate prediction outcomes and assist proactive mental health treatments. Duell et al. [174] proposed Counterfactual-IG for medical records, which provide counterfactual explanations to help clinicians make better decisions and improve patient outcomes.

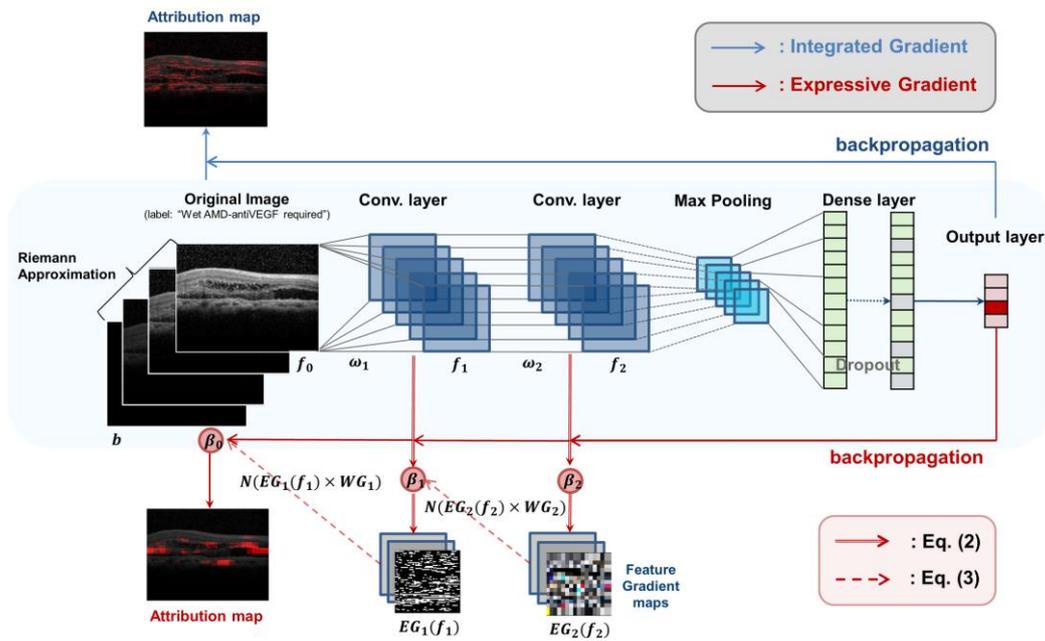

**Figure 9.** An application of IG for weakly supervised lesion localization for age-related macular degeneration detection using optical coherence tomography images [234].

Acharya et al. [175] used a Jump Knowledge Based Cell Graph Neural Network and IG to predict TB from lung tissue pictures, revealing insights into tuberculosis-related traits. Dolci et al. [176] investigated bias in XAI attribution approaches, including IG, to eliminate biases that impair AI fairness and dependability in extended reality and neural engineering applications. Lentzen et al. [177] created a transformer-based model for forecasting severe COVID-19 development that uses IG to understand claims data, therefore enabling transparent and dependable AI-driven crisis management in healthcare. Abeyagunasekera et al. [81] developed the LISA framework to unify XAI approaches in medical imaging, using IG to improve diagnostic accuracy and confidence in AI-assisted medical image processing.

Researchers in [178] introduced a hybrid Integrated Gradient-based model for tracking things in video frames and demonstrated its use in video analytics to improve accuracy and dependability. Lamba and Rani [179] used IG to diagnose brain tumor problems using DL models, improving the interpretability of AI-based neuroimaging diagnostics in clinical settings. These studies demonstrate how IG can improve the transparency, interpretability, and reliability of AI models in a variety of critical medical domains, including ophthalmology and neurology, as well as infectious diseases, mental health, and oncology diagnostics. Figure 9 illustrates the application of integrated gradients algorithm localization for age-related macular degeneration detection using optical coherence tomography images.

### 4.5. Partial Dependence Plots (PDP)

The use of PDP is gaining popularity in research concerning the diagnosis and management of various health conditions, providing significant insights into the advantages of using XAI in the health sector. This section discusses the application of PDP across different medical specialties to enhance predictive modeling and interpretability.

In the context of cancer research, a study used a fully nonparametric random survival forest (RSF) model incorporating PDP to assess 3-year survival rates and identify prognostic factors [180]. The results indicated that older patients, lack of insurance, and advanced-stage cancer had poorer prognoses. The RSF model demonstrated superior discriminative capacity and predictive accuracy. [181] conducted a multicentre study that aimed to develop and validate ML algorithms for predicting lymph node metastasis in early gastric cancer. Using algorithms such as XGBoost alongside PDPs, the study identified higher metastasis rates among young women with specific tumor characteristics. XGBoost provided significant predictive performance, and the authors converted the model into an accessible web tool for clinicians.

In geriatric studies, research on elderly patients with osteoarthritis suggested that co-occurring cardiovascular and gastrointestinal disorders could be predicted using RF models with PDP [183]. Strong predictors included age, cardiac arrhythmia, and extended use of corticosteroids. PDP effectively demonstrated the increased risk associated with these factors, highlighting the method's ability to enhance the interpretability of complex models.

In a stroke study, researchers used the LightGBM model with SHAP and PDP to predict early neurological deterioration [153]. The model's performance was highly accurate, and PDPs individualized feature effects, improving the model's interpretability. Another study employed RF regression and PDP to predict the 90-day home time for stroke patients post-endovascular thrombectomy [143]. Factors such as atrial fibrillation, hypertension, and diabetes were associated with lower home time. PDPs showed comparable outcomes for patients with low and high ASPECTS, emphasizing the model's robustness.

A study on depression classification used XAI with PDP, where the XGBoost algorithm achieved high accuracy [184]. PDPs, along with SHAP, provided complementary insights into the relative importance of different features, enhancing the classification of depression.

Research investigating the association between heart disease and socioeconomic factors found that low income significantly increased the odds of heart disease, particularly in men below 70 and women below 80 [185]. The authors used PDP to demonstrate correlations between the risk of heart disease and factors such as age, sex, smoking, income, and education level. PDP supported the robustness of these relationships, independent of specific modeling assumptions. Using RF models and PDP, [186] examined gender-specific cardiovascular risk factors in a large cohort. Results indicated differences between men and women, with higher cholesterol levels posing greater risks for men and ageing and increased waist circumference posing higher risks for women. In a study on chronic heart failure patients, SHAP values and PDP were used to interpret ML models predicting 3-year all-cause mortality [187]. Key predictors included age, cardiac function, and hospitalization. PDPs offered clear individual risk factors, aiding clinicians in their decision-making.

Similarly, [188] employed RSF models with PDP to predict post-mitral valve replacement mortality in severe degenerative mitral regurgitation patients. Significant predictors included age, ventricular dimensions, and hemoglobin levels. PDPs helped develop effective stratification scores, enabling clinicians to make informed decisions. A longitudinal study employed an RF model with PDP to identify early predictors of Alzheimer's disease [189]. Key predictors included daily life functioning, hippocampal volume, and verbal memory, with subtle changes reported by family members highlighted by PDPs, providing valuable insights for early diagnosis.

Using MRI and clinical data, [139] developed an explainable ML model (MUQUBIA) to distinguish different neurodegenerative dementias. SHAP and PDP analyzed the data, supporting the critical features of classification and improving the accuracy and interpretability of decision-making. [190] used SHAP and PDP, alongside gradient boost models, to identify predictors for extubating failure in critically ill COVID-19 patients. Key predictors included neurological status, inflammatory parameters, and ventilatory settings. PDP contributed critical values for the probability of extubating failure, enhancing clinical decision-making.

In the context of COVID-19, an automated ML model predicted mortality, identifying key predictors such as oxygen levels and blood pressure [191]. PDP refined the model, underscoring the most significant predictors for mortality among COVID-19 patients.

These studies showcase the influence of PDPs in improving the interpretability of ML models, enabling better diagnosis and management of various medical conditions. PDPs facilitate informed decision-making by visualizing the marginal effects of predictors and providing valuable insights into the relationships between features and outcomes.

*4.6. Permutation Feature Importance (PFI)*

This section focuses on the integration of PFI in human health monitoring. In [45], an explainable prediction model for HF survival is proposed and presented. The proposed idea is based on a dual approach, first, addressing the survival prediction through a survival analysis and a classical ML classification problem. The PFI technique has been used among other explainability metrics to detect the best-balanced model in terms of prediction and explainability.

In [46], several ML models have been used to provide an accurate and highly explainable ensemble model to successfully predict the risk of diabetes. It was demonstrated that the predictions based on the weighted ensemble model (based on the two best-performing algorithms) are significantly better than the individual algorithms. The PFI technique is used along with other techniques to help physicians understand the model predictions.

In [47], the authors tried to shed light on the acceptability of XAI methods from the domain experts' (12 oncologists) perspective. The goal of the research was to verify whether XAI techniques accomplish the explanation needs of domain experts. In this regard, the PFI and other techniques were used and assessed, and a question was asked to the domain experts whether they were able to identify the least important features influencing the results presented to them or not.

In [91], the authors present an ML approach based on an RF model for an accurate prediction of Upper Limb (UL) functional recovery scores at discharge after inpatient stroke rehabilitation. PFI along with other techniques allowed interpretation of the ML outcomes and to identify those variables having a more prominent role in the prediction, reporting a consensus across the top features identified by all the chosen methods. Figure 10 gives the general workflow illustrating this study. From a clinical perspective, the findings of this work may improve the management of sub-acute stroke patients in a rehabilitation unit.

In [166], the authors have designed an AI-based COVID-19 diagnostic method using Mass Spectrometry (MS) data and leveraged XAI to explain the decision-making process on a local (per-sample) and global (all samples) basis. Eight ML models with five feature engineering techniques using a five-fold stratified cross-validation were tested. PFI along with other techniques was used to rank the 10 ($R_0$, $R_1$, $R_2$, $R_3$, $R_4$, $R_5$, $R_6$, $R_7$, $R_8$, and $R_9$) features tested.

In [117] the researchers have introduced an easily interpretable approach for heart stroke prediction using XAI techniques. This approach incorporates pivotal techniques such as resampling, data leakage prevention, feature selection, and emphasizing the model's understandability for healthcare practitioners. Several AI models have been used which are: RF, Logistic Regression (LR), SVM, eXtreme Gradient Boosting (XGBoost), and ANN. The PFI has been used to determine the importance of each feature. For LR the most important features ranked from the most important to the lowest are age, residence type, smoking status, gender, and work type. PFI has been used to determine the importance of the 12 considered features and the importance changes based on the AI model used.

In [192] the authors have introduced a new permutation-based feature importance test called PermFIT to estimate the importance of features. The proposed method has been applied to the applications to TCGA kidney cancer data and HITChip Atlas BMI data and 3 AI models were used namely DNN, RF, and SVM. The proposed approach, PermFIT, has been proven to be a computationally efficient permutation-based feature importance and it has demonstrated its practical usage in identifying important biomarkers and boosting AI model prediction performance. The application of the new PermFIT has identified the most important features based on the model used.

*4.7. Counterfactual Explanations (CE)*

Counterfactual explanations provide causal inferences in AI systems by suggesting what could be different to change the outcome of the system. Hence, they provide the highest level of explainability or interpretability by elucidating the reasons behind AI systems' decisions. Advancements in big data analytics, high-performance computing, and ML promise to revolutionize precision medicine allowing for the identification of disease risks, preventive measures, improved diagnosis, and personalized treatment. Hence data-driven models are being considered for critical decisions. The black-box nature of AI models means that they are often not trusted by both practitioners and patients. Therefore, traditional XAI techniques are often employed for interpretability. While most of these models can give explanations by identifying the most important parameters and their correlation with observed model decisions, they do not give causal inferences [193]. This could lead to unintended consequences. Hence, there is a recent interest in the use of counterfactuals to obtain actionable explanations. CF has been considered in various applications across healthcare including diabetes Miletus [194], [195], [196], cancer [194], [195], [196], [174], [197], [198], posture deficit [199], and Alzheimer's disease [143]. These applications are summarized in Table CF.

In response to concerns from the ML community and regulatory requirements, a unified tool addressing issues of explainability, fairness, and robustness was developed in [195]. CERTIFAI, which is based on a genetic algorithm, was proposed to generate counterfactual explanations for actionable recourse. This tool improved the realism of constraints by muting unchangeable features and imposing restrictions on both feature range and the number of generated explanations. CERTIFAI was applied to the Pima Indian dataset using AI models such as Inception-v3, ResNet-50, and MobileNet for diabetes mellitus classification. The framework is model-agnostic and flexible, ensuring that explainability is achieved through the development of feasible counterfactual explanations. The features of CERTFAI allowed for the generation of counterfactuals for nonlinear models while solving the problem of infeasibility using earlier methods. The results obtained with CERTIFAI were further extended in [196].

The Adaptive Feature Weight Genetic Explanation (AFWGE) [196] also employs a custom genetic algorithm to obtain counterfactual explanations, utilizing adaptive feature weights to enhance performance. This method was applied to four different datasets using a multi-layered perceptron (MLP) and benchmarked against CERTIFAI. AFWGE produced more effective counterfactuals with superior proximity, sparsity, plausibility, and actionability. Beyond generating improved counterfactuals, AFWGE provides a robust framework for assessing feature importance in ML models. The performance of AFWGE was compared to CERTIFAI using four datasets, including two healthcare-related ones – the Pima Indian Diabetes dataset and the Breast cancer Wisconsin (Diagnostic) dataset. AFWGE demonstrated improved performance over CERTIFAI in terms of the mean distance between the original instance and generated counterfactuals, as well as the number of feature changes. In the diabetes study, both CERTIFAI and AFWGE identified BMI and glucose levels as the most important features for predicting diabetes risk, with skin thickness being less important.

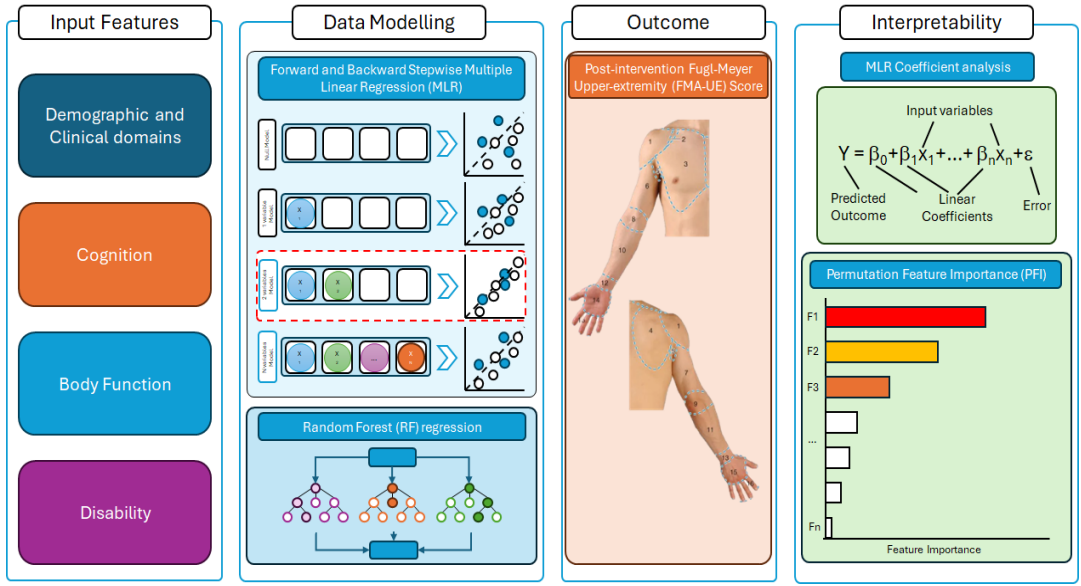

**Figure 10.** An example of using the PFI in one study to determine an accurate prediction of upper limb (UL) functional recovery scores at discharge after inpatient stroke rehabilitation.

Advances in medicine and imaging allow for the use of Neoadjuvant Systemic Therapy (NST) in the treatment of early-stage breast cancer. However, the prediction of pathological complete response (pCR) to NST is not straightforward, and ML tools are being suggested to alleviate this challenge [200], [201], [202], [203]. However, the non-transparency of these methods limits their clinical applicability. While earlier explainability methods capture the correlation between features and responses, counterfactual methods such as [197] are proposed to find causal relationships. In [197], causal relationships between ICM features and patients' pathological response to NST are predicted before treatment. This was achieved by initially filtering the features that do not contribute to treatment response prediction using Shapley values before carrying out counterfactual analysis using DiCE [204]. The cause-and-effect relationships between breast MR imaging phenotypes (variance of IV uptake, Mean washin slope, margin fluctuation, tumor solidity), molecular features (PR, ER subtypes), and pathologic response to NST were identified. However, DiCE is computationally heavy, and the selection of only 10 features out of the original 536 features significantly limited the feature space.

In [194], the work in [197] was extended to generate more sparse, feasible, plausible, and diverse counterfactuals at a lower computational cost. The proposed Sparse CounteRGAN (SCGAN) improved on CGAN by producing sparse but diverse counterfactuals while ensuring plausibility. The Pima Indians Diabetes dataset and Ionosphere dataset were used as benchmarks while the Breast Cancer DCE-MRI dataset was used as a case study. In both the diabetes and breast cancer studies, SCGAN generated sparse and diverse counterfactual instances that were quite plausible and feasible. The counterfactuals showed a causal relationship between two features (Glucose and BMI) and three ICM features (PR status, HER2, and variance of uptake) in the diabetes and cancer treatment cases respectively. The results help guide oncologists in the treatment of cancer patients who find it difficult to achieve pathologic complete response (pCR) by providing patient-specific features while eliminating confounding factors making it a valuable tool for understanding the causal relationships between ICM features and treatment response. This approach helped identify causal relationships between imaging phenotypes, clinical history, molecular features, and pathologic response to NST. For illustration purposes, the SCGAN method is shown in Figure 11. Radiomic features from the MR image are used to generate counterfactual instances.

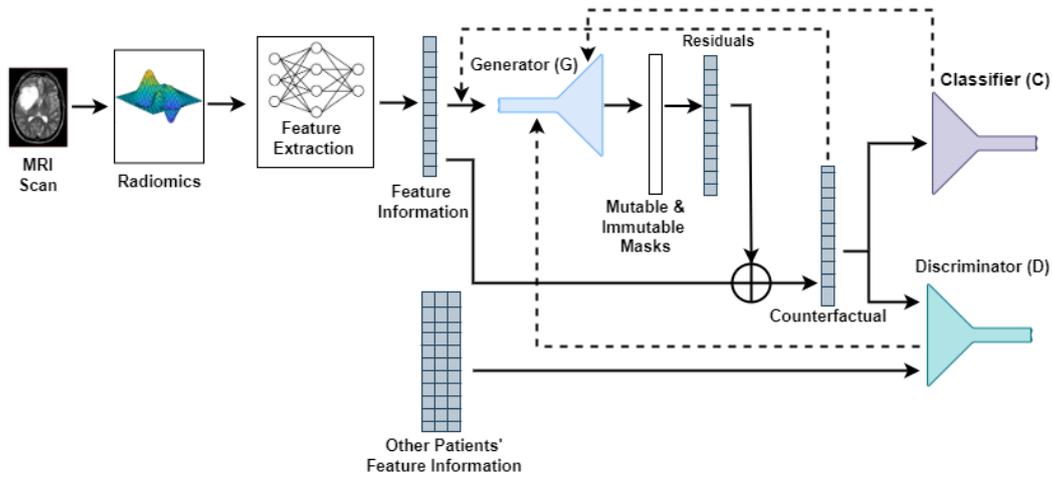

**Figure 11.** The SCGAN method on an example of breast cancer MR image using Counterfactual Explanations.

In [198], CLARUS was developed to enhance the understanding of graph neural network (GNN) predictions. The framework allows users to pose counterfactual questions based on initial insights from XAI methods such as GNNExplainer, IG, and Saliency. This enables the retraining of the GNN after manual manipulation using counterfactual questions. The Kidney Renal Clear Cell Carcinoma (KIRC) dataset from the Cancer Genome Atlas (TCGA) database [205], [1206] was used to demonstrate the functionality of CLARUS in binary classification tasks. This paper emphasizes the use of explainability and causality to build trust in AI models for medical applications. Importantly, human-in-the-loop interactions enhanced causality and explainability for kidney cancer prediction scores using the considered networks.

Traditional methods for assessing postural deficits via visual inspection of back contours are subjective and prone to errors. This has prompted the development of machine learning-based methods, particularly XAI techniques for transparency. In [199], the subjects' spine characteristics were used to calculate the features of kyphosis index (KI), flèche cervicale (FC), and flèche lombaire (FL), which were normalized using the subjects' trunk heights. DiCE was utilized to find causal relationships with postural defects. The results showed causality relationships as follows: KI% and flèche lombaire (FL%) in hyperlordosis, and KI% and flèche cervicale (FC%) in hypokyphosis.

In [174], Counterfactual Integrated-Gradients (CF-IG) was proposed as a new technique for causal analysis using counterfactuals. This method was used to evaluate patients by generating hypotheses based on existing data without the need for physical intervention. The symbiosis between CF and IG provides a robust framework for causal analysis while retaining essential XAI principles.

Causal inferences are paramount for critical decisions in human health applications. The use of counterfactual explanations in explainable AI for healthcare is a growing area of research. Therefore, most of the work available in the literature focuses on developing methods and frameworks for generating counterfactuals. Most of the applications to healthcare data are used to test the developed methods in conjunction with other non-medical data. Healthcare applications are mostly not the primary objective of these studies. Hence, the area is relatively virgin with a lot of opportunities for research. Moreover, only a few diseases have so far been studied using CF techniques. Perhaps, this could be because a lot of diseases have already been studied and understood by experts. However, with the availability of massive amounts of data and advancements in computing and AI techniques, we believe that the method of CF has great potential to revolutionize the ethical application of AI in human healthcare and related sectors.

*4.8. Saliency Maps (SM)*

The use of saliency maps in XAI methods is very common in the field of disease identification. [169] presents a comparison of seven SM-based XAI methods to diagnose epiretinal membrane, a type of ocular disease that occurs in the human retina. Here, CNN is used as a DL tool to screen the 2-D images of the retina using optical coherence tomography. A study on Parkinson's disease patients is discussed in [207] where brain images have been trained with 3-D CNN. With the use of SM in the applied XAI, it has been found that the orbital-frontal cortex, frontotemporal regions, and multiple deep gray matter structures are the significant features of the study. A similar study is presented in [208] on Parkinson's disease patients where SM identifies that diffusion tensor imaging data, like fractional anisotropy, is more effective than T1-weighted data for classification purposes. Article [209] has presented a study to diagnose Myocardial infarction (MI) rapidly and efficiently, using five explainable DL techniques. Among them, InceptionResNetV2 achieved the best accuracy but as per the Grad-Cam results, ResNet50V2 became the most trusted one. In [210], a saliency map was used on a pre-operative video-based echocardiogram dataset. The saliency map was used in the articles [211], [212], [213] in identifying Alzheimer's disease (AD), a progressive neurodegenerative disorder. With four CNN methods, two ensemble models are formed in [211] for the detection of AD which results in up to 96% accuracy (see figure 12). Both the Grad-CAM and saliency map were used as XAI tools in this work to recognize the illness symptoms. The achieved results can guide clinicians in making insightful decisions by analyzing the neural regions. Article [212] proposed a visual saliency detection method to identify the abnormality more effectively. This paper used an elliptical local binary pattern descriptor for MRI image characterization. Top-down and bottom-up saliency maps were used for domain knowledge and image features, respectively. Paper [213] used biomarkers in EEG signals to study AD where power modulation spectrogram inputs are trained to detect optimal patches in a data-driven strategy. The use of a saliency map in the DL model not only helped in identifying the AS but also the severity level and cause. A saliency map was proposed to detect the skin disease in [214], with the help of ConvNets. The paper proposed a new XAI model for the purpose where U-Net was used for segmentation and MobileNet V2 was used for segmentation. In [215], SM was used to identify two eye diseases namely diabetic retinopathy (DR) and neovascular age-related macular degeneration (nAMD). In this work, an ensemble model of three different network architectures was used to detect these diseases using retinal fundus images and optical coherence tomography scans, respectively. It has been found that the efficacy of SM depends upon the DNN architecture & explanation method. An SM-based analysis was presented in [216] on schizophrenia patients. After analyzing the saliency-based static images, it has been found that the Visual-oculomotor properties were affected in schizophrenia, but oculomotor properties were not. Besides, Visual salience affected the gaze of such patients. It has also been observed that the Orientation salience is correlated with the scores of cognitive tests and visual oculomotor characteristics. Article [217] used the saliency analysis to compare between the localization of breast lesions in mammogram screenings against the areas of interest from AI systems. The key feature of using SM was that it generates 'heatmaps' in the presence of suspicious regions or lesions. Thus, the work explained a detailed process of detecting malignancies with four different deep-learning models. Among them, END2END and DMV-CNN provided better performance than others.

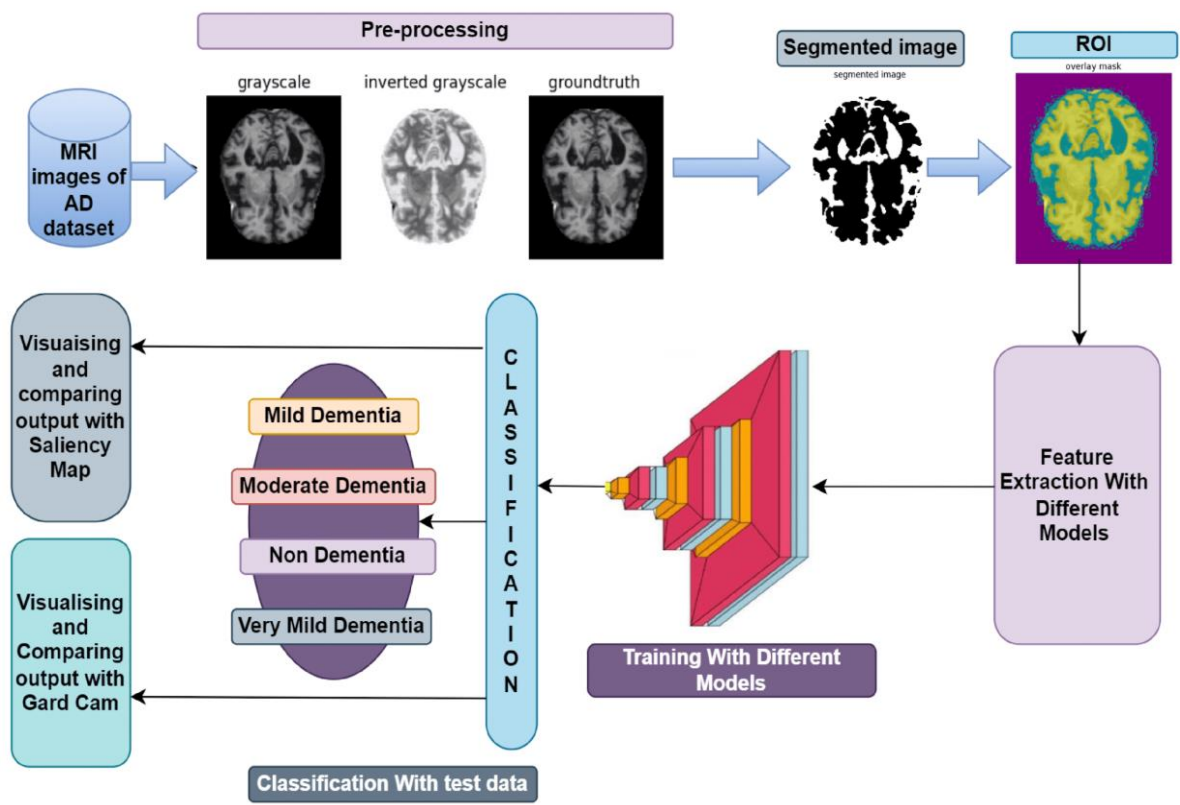

**Figure 12.** Saliency Maps for AD detection by diagnosing MRI images [211].

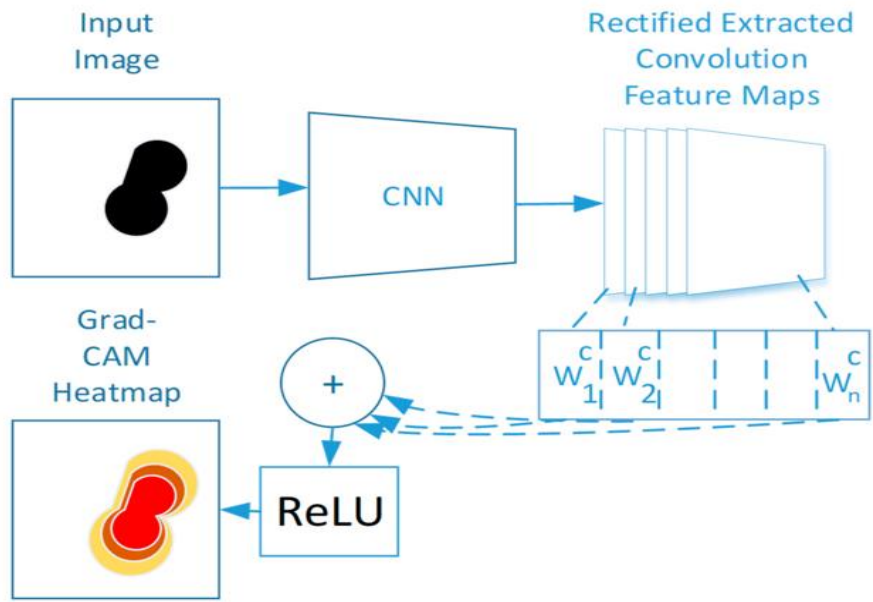

**Figure 13.** Grad-CAM block diagram Flowchart and components [223].

*4.9. Gradient-weighted Class Activation Mapping (Grad-CAM)*

Grad-CAM is an interpretability technique created to visualize the areas of an input image that have the most influence on a specific prediction, hence increasing the explainability of CNNs as shown in Figure 13. Grad-CAM accomplishes this by creating a rough localization map that highlights significant areas of the image employing the gradients of any target class flowing into the final convolutional layer. This facilitates comprehension of the regions of the image that the model is concentrating on, offering insights into CNN's decision-making process. Grad-CAM is especially useful for situations where interpretability and confidence in model predictions are critical, such as medical imaging [218], [103].

The authors in [103] proposed XAI which has become increasingly important in the field of medical imaging, particularly for critical jobs such as CT images for Covid-19 diagnosis. When it comes to bias identification, XAI approaches can be applied to CT-scan classifiers to discover biases, particularly when unbalanced or demographic characteristics are considered. XAI can offer both individual prediction level and overall model behavior, which may be used to detect and comprehend biases based on the local and global explanations. Feature importance analysis using XAI clarifies which features the model depends on, allowing one to determine whether biases are due to important causes or not. Furthermore, XAI improves transparency and patient and clinician confidence in the AI's diagnostic process by providing an explanation of model decisions. XAI insights guide the removal of unnecessary or deceptive characteristics from models, removing bias and increasing accuracy and fairness. By making it easier to identify and resolve discriminatory patterns in diagnostic tools, the application of XAI approaches helps to ensure algorithmic fairness and, eventually, improve equitable healthcare outcomes. The results of XAI highlight the significance of strong diverse training data. They highlight the potential for specific datasets to unintentionally introduce bias into the model, guiding future attempts to gather data to be more representative and inclusive. These results encourage cooperation between medical practitioners and AI specialists, resulting in a multidisciplinary approach to the development and improvement of fair, accurate, and explicable diagnostic tools. In general, using XAI for bias identification in Covid-19 CT-scan classifiers is a critical step towards developing more precise, equitable, and comprehensible diagnostic instruments for medical imaging.

According to [219], XAI on Covid-19 classification from images from chest X- has demonstrated how DL models such as CNNs have a high diagnostic accuracy for the disease. By emphasizing important areas of the X-ray image that are crucial for identification, XAI addresses Grad-CAM and saliency maps to enhance model interpretability and lower bias. Since it contributes to the diagnosis process and prediction confidence more understandable, this transparency boosts clinician trust and adoption. Additionally, by offering early screenings, XAI reduces the load for radiologists and facilitates iterative model improvement by locating and fixing flaws. The obtained findings additionally assist in validating and regulatory compliance, functioning as a resource for education regarding the integration of AI tools into clinical workflows and, eventually, improving patient experiences and clinical decision-making. The authors in [220] demonstrated that Covid-19 classification based on Grad-CAM visualization and fusion-assisted deep Bayesian optimization utilizing chest X-ray images demonstrates that deep Bayesian optimization combined with Grad-CAM significantly enhances model interpretability and accuracy. DL models are further developed through the application of Bayesian optimization, which results in superior Covid-19 detection performance. Transparency and clinician trust are strengthened by Grad-CAM visualization, providing insights by emphasizing the crucial areas of the X-ray images utilized for predictions. By combining several features and model outputs, the fusion-assisted method produces an identification system that is more reliable and robust. In addition to improving diagnostic accuracy, this integrated framework provides thorough visual explanations, which facilitate better comprehension and validation of AI-driven diagnosis in healthcare environments.

Table 1. Summary of some selected articles on different XAI techniques

| Ref | Citation | ML | Disease | Data Type | XAI | Result of XAI |
|---|---|---|---|---|---|---|
| [83] | 423 | DNNs | Motor imagery brain computer interface | EEG | LRP | LRP produces neurophysiologically highly plausible explanations of how a DNN reaches a decision. |
| [22] | 248 | CNN | Gait biometrics | Motion capture, GRF, EMG | LRP | Gait events to identify individuals |
| [195] | 60 | DNNs | Diagnostic model | Brain tumor images | LIME | Insights into neuroimaging data for accurate diagnosis and treatment planning. |
| [137] | 225 | RF | Alzheimer's disease | Medical Data | SHAP | Prediction performance compared to other models. |
| [210] | 38 | 3-D CNN | Heart disease | Video-based ECG data | SHAP | Predict the onset of disease using echocardiography ML |
| [81] | 21 | DNNs | Alzheimer's disease | Neuro imaging data | IG | Insights into disease progression mechanisms. |
| [187] | 134 | Statistical Analysis | Chronic heart failure | Demographic and medical information | PDPs | Explanation of individual risk prediction and give doctors an intuitive knowledge of the functions of important model components. |
| [191] | 78 | Statistical Analysis | Surviving a SARS-CoV-2 infection. | Demographic and medical information | PDPs | Learning-based clinical decision support tools. |
| [195] | 132 | CNN | Diabetes Miletus | Medical Data | CF | Performance analysis diabetes prediction |
| [215] | 37 | DNNs | Ophthalmology | Retinal images | SM | Performance analysis |
| [192] | 84 | DNNs | Cancer | Reverse Phase Protein Arrays | PFI | Cancer's most important feature among Age, nationality, and microbiomes identified. |
| [46] | 50 | DNNs | Diabetes Mellitus | Attributes of diabetes | PFI | Diabetes's most important feature among Glucose, pregnancy, age, and BMI, blood pressure, skin thickness |

As mentioned in [221], Grad-CAM when used with three-dimensional (3D) CNN can significantly enhance the early detection of lung cancer. Important findings for this study show that this method not only increases the precision of detecting malignant nodules but also offers a visual representation of the decision-making process, facilitating comprehension of the areas that contribute most to the categorization. This approach helps radiologists make more informed decisions about early lung cancer diagnosis and improves their ability to interpret the results. The integration of DL models Grad-CAM greatly enhances the diagnosis and localization of brain tumors, according to [222]. Its obtained results demonstrate that by highlighting critical locations in medical images, this combination not only increases the sensitivity and accuracy of tumor diagnosis but also offers useful visual explanations. This helps doctors make more accurate and confident diagnoses by improving the interpretability and reliability of detection outcomes. The authors in [223] concluded that Grad-CAM is an extremely helpful instrument for enhancing the accuracy and understanding of diagnostic assessments rendered in various kinds of medical imaging environments. Their findings indicate that Grad-CAM improves physicians' understanding and trust in the diagnostic process by providing accurate visual explanations by emphasizing spots in the images that are essential to the models' predictions. More accurate and trustworthy medical diagnoses are rendered feasible by this tool's assistance in identifying important regions.

The authors in [224] realized that significant factors influencing joint contact forces at the knee during walking can be efficiently recognized and depicted by combining motion analysis techniques with Grad-CAM. The results demonstrate that Grad-CAM is effective in identifying key motion patterns and biomechanical elements that result in knee joint stress, which increases the understanding of the connection between knee health and gait mechanics. This method assists in establishing specific therapeutic methods by improving the ability to diagnose and perhaps alleviate diseases associated with aberrant knee joint forces. On the other side, DNNs can diagnose skin disorders from medical photographs at a level of performance that is comparable to that of dermatologists. The main results indicate that DNNs consistently differentiate between several kinds of skin lesions, including benign and malignant cases, typically matching or even exceeding the diagnostic precision of skilled dermatologists. This illustrates how DNNs have the potential to be useful assistive tools in dermatology, increasing diagnostic efficiency and accuracy while providing valuable help to healthcare providers [225].

The authors in [226] proposed a framework that specifically finds and recognizes brain atrophy, and significantly enhances the process of testing for Alzheimer's disease. The model's capacity to recognize intricate patterns in MRI scans has been enhanced by the addition of residual connections and self-attention mechanisms. Additionally, by emphasizing pertinent areas related to the illness, the explainability of the model—which is rendered feasible by these mechanisms—allows for transparent decision-making. As a result, the model is more beneficial for clinical applications by increasing the accuracy of Alzheimer's diagnosis and the interpretability of its forecasts. The authors in [227] demonstrated how electrocardiogram (ECG) data of elderly post-stroke patients can be effectively analyzed using a one-dimensional (1D) CNN combined with Grad-CAM. The findings reveal that employing this approach improves the precision of identifying irregularities in ECG rhythms that point to post-stroke disorders. By emphasizing particular ECG signal characteristics that support the model's predictions, Grad-CAM offers useful visual insights that help doctors comprehend and validate the results. This approach presents a potentially useful tool for enhancing cardiac problem monitoring and diagnosis in older post-stroke patients.

According to [228], the accuracy of Parkinson's disease diagnosis is much increased when utilizing a 3D ResNet (3-dimensional Residual Network) model. The model highlights the frontal lobe's role in differentiating between people in a healthy state and those suffering from Parkinson's disease, in line with key findings. 3D ResNet attains greater diagnostic precision by concentrating on the anatomical and functional alterations in the frontal lobe. This highlights the crucial role that this area of the brain plays in the pathophysiology of the disease, offering further insight into the diagnosis of Parkinson's disease and maybe guiding the development of more focused treatment approaches. The use of a parameter-weighted structural connectome matrix in conjunction with DL techniques significantly improves the capacity to diagnose Parkinson's disease as stated in [229]. The results show that this method helps identify particular neural circuit problems associated with Parkinson's disease and enhances diagnostic accuracy by accurately capturing the intricate structural connection patterns associated with the disease. The parameter-weighted matrix reveals important neuronal connections and pathways that are broken in patients, presenting a viable avenue for research into targeted therapies as well as a thorough understanding of the brain foundations of the disease. Furthermore, wearable technology in conjunction with an improved DL model can reliably and continuously monitor and detect Parkinson's disease in daily life [230]. The main findings demonstrate that this strategy, which makes utilization of wearable sensor data, not only provides high accuracy in identifying symptoms associated with Parkinson's disease, but also enables interpretability by highlighting the features of the sensor data that are most significant in making detection decisions. This makes the system easier to understand and more trustworthy when used in real-world scenarios for Parkinson's disease early diagnosis, individualized treatment, and ongoing patient monitoring.

According to [231], the identification of late mechanical activation in the heart is much improved when a multitasking learning strategy is used. The model is trained to simultaneously optimize several related goals, which enhances its capacity to recognize the intricate and nuanced patterns linked to heart tissue mechanical activation delays. By utilizing shared representations between tasks, the multitask learning framework outperforms single-task models in terms of diagnostic performance. The incorporation of Grad-CAM for boosting interpretability is not explicitly addressed, despite the fact that the study greatly improves detection skills. For the fast and precise identification and treatment of heart-related disorders, this technique offers more robust and accurate detection capabilities, which can be critical. The research [232] indicates that myocardial infarction (MI) diagnosis and interpretability from electrocardiogram (ECG) signals are significantly enhanced by merging DL models with Grad-CAM. The main findings of this research demonstrate that DL models identify MI with high accuracy, and the application of Grad-CAM highlights the crucial ECG signal segments that support the model's predictions and offer understandable visual explanations. This enhances the diagnostic performance of the model and helps physicians interpret and confirm the findings, building their confidence in the automated detection system and may lead to better patient outcomes.

## 5. Discussion

Over the past decade, ML and XAI have significantly advanced in the field of disease diagnosis. This review commenced with targeted research questions, aiming to address them through an analysis of existing references. The findings indicate that CNNs are particularly notable, as superior when combined with XAI algorithms to understand disease from the AI perspective due to their strong capabilities in handling both image and tabular data [55], [76], [77], [78], [79], [80], [81], [68], [106], [109], [110], [115], [169], [213], [52], [103], [219], [221], [227].

several factors and could be achieved considering tradeoffs.

Table 2: Comparison between the XAI tools

| XAI Method | Type | Strength | Weakness |
|---|---|---|---|
| LIME | Model-agnostic, Local | Flexible, capable of working with any black-box model, and good for individual predictions. | Results can be unstable, depending on the perturbation method and local model. |
| SHAP | Model-agnostic (with variants), local & global | Consistent, theoretically sound, works globally and locally. | Computationally expensive, especially for large models and datasets |
| LRP | Model specific | Visual relevance representation is useful for DL models & noise-mixed models. Uses backpropagation. | Complex implementation, sensitive to model architecture |
| PDP | Model-agnostic, Global | Simple, intuitive visualization of feature effects. | It may not capture complex feature interactions and can be misleading if features are correlated |
| PFI | Model-agnostic, Global | Straightforward, easy to implement | Computationally intensive, less effective with high feature correlation |
| CF | Model-agnostic, Local | Intuitive provides actionable insights | Computationally expensive, not always feasible or meaningful |
| IG | Model-specific | Consistent measure of feature importance, works well with neural networks | Chooses a baseline, less intuitive compared to some other methods |
| CM | Model-specific, Local | Useful for visualizing influential parts of the input | Can be noisy, sensitive to perturbations, less clear actionable insights |

As revealed through this comprehensive literature review, the integration of XAI in human health monitoring showcases a dynamic interplay of innovative applications, significant challenges, and promising future directions. XAI has made notable strides in enhancing healthcare delivery, particularly in medical imaging [110], [111], [215], [86], [87], [81], [212], [66], [68], [69], 100], [101], [124], [226], [202]; where it demonstrates remarkable capabilities in enhancing diagnostic accuracy and early disease detection. This is further complemented by its integration into wearable health

monitoring devices [230], [56], [90], [55], [71], [73], [74], [227], [232], [59], [119], [83], [109], [213], enabling continuous, real-time tracking of vital health parameters. These advancements are pivotal in managing chronic diseases like Parkinson's, stroke, depression, cancer, heart disease, and Alzheimer's disease more effectively.

Table 1 presents a summary of different XAI techniques, with some highly cited references discussed in the previous sections. Various XAI methods are tried for various diseases, and the success rate varies with different determining factors. cation depends on

*5.1. Comparison between the XAI methodologies*

It's difficult to comment on which method is superior to others as it depends upon various factors. Some models like SHAP, LIME, and CF focus on local explanations, whereas PFI, PDP, and global SHAP models rely on overall model behavior. Some XAI tools are model specific whereas others are open to use with any ML models. In terms of complexity, PDP and PFI are simple in visualization, but the features are difficult to control. On the other hand, LIME and SHAP models are more interpretable in nature if handled carefully; Table 2 presents a fair comparison among the XAI methods for future reference. CF, on its part, tends to give causal relationships and intuitive actionable explanations, but at a high computational cost. Hence, choosing an XAI technique for any application depends on several factors and could be achieved considering tradeoffs.

*5.2. Challenges in using XAI in healthcare*

Researchers and practitioners can work towards a more comprehensive and holistic understanding of AI model behavior by leveraging the strengths of various XAI methods, including the well-established LRP approach and the promising counterfactual explanations. This, in turn, can lead to increased trust, transparency, and responsible deployment of AI systems across a wide range of applications, from healthcare and finance to policymaking and beyond.

In the following sections, we highlight the key challenges and important insights of using XAI methodologies in healthcare management.

**5.2.1. Human interaction capability**

Simply explaining how a ML model works is not sufficient for effective health monitoring and disease detection. Clinicians need to deeply understand the model's decision-making process to trust and verify its diagnoses. Thus, making the systems interactive is a challenge in this addressed field. The promising research direction in this context is the development of interactive explanation systems tailored to healthcare. These systems would allow clinicians to actively engage with the model, moving beyond passive information consumption. For example, clinicians should be able to ask follow-up questions, explore alternative diagnoses, visualize how the model arrived at a specific conclusion, or adjust model parameters to see how different inputs affect the diagnosis [235]. This interactive, iterative approach would help clinicians develop a more comprehensive understanding of the model's reasoning, ensuring they can trust its outputs and make informed decisions. Furthermore, these interactive explanation systems could be customized to fit the specific needs of different medical specialties and expertise levels.

**5.2.2 Economic Prospective**

The technical challenge of balancing accuracy with interpretability also impacts the cost of developing and implementing XAI in healthcare systems [236]. Interestingly, while much of the literature focuses on XAI's technical and ethical aspects, the economic implications are less explored but equally important. Understanding the economic perspective of XAI is vital for addressing challenges such as the costs associated with achieving explainability, the impact of interpretability on clinical decision-making, and the potential market evolution of XAI-driven healthcare tools. Encouraging economic analyses of XAI in healthcare could help estimate the costs and benefits more accurately, address concerns about proprietary algorithms and trade secrets, and provide insights into the future market landscape for XAI-enabled medical technologies.

**5.2.3 Ethical point of view**

Ethical and privacy concerns are at the forefront [237], especially considering the 'black box' nature of some AI algorithms, which raises significant ethical dilemmas in decision-making. The increasing adoption of wearable technologies and remote monitoring devices amplifies these privacy concerns, emphasizing the need for stringent data security measures to sustain trust in AI-based systems. Improving the interpretability and transparency of AI systems is critical as it enhances data integration across different platforms and develops sophisticated disease prediction and management models. Moreover, the dynamic nature of AI technology necessitates adaptive regulatory frameworks that can ensure the safe, effective, and ethical use of AI in health monitoring. These frameworks should be capable of evolving with technological advancements while protecting patient interests. Addressing these regulatory and ethical aspects and focusing on data security and patient privacy is crucial for maintaining public trust and compliance with legal standards.

**5.2.4 Selectiveness of XAI techniques**

Selecting the most appropriate XAI analyzer for a ML model is a significant challenge, particularly in medical diagnostics with high stakes. ML-based diagnostic models directly impact human lives, and every instance of misdiagnosis poses a potential threat to patient health [237], [238]. Therefore, ensuring that the XAI methods employed are accurate and reliable in highlighting the key features used by these models for classification is critical. Despite the crucial nature of this task, many studies have not adequately evaluated the effectiveness of XAI methods, especially in the presence of noise. Noise in data can obscure the true features that should be used for classification, leading to misinterpretations and, consequently, misdiagnoses. Therefore, assessing whether the selected XAI method can accurately capture the essential features in noisy environments is paramount.

**5.2.5 XAI Ground-Truth and Data**

The successful deployment of Explainable Artificial Intelligence (XAI) in medical diagnostics is a multifaceted process that requires effective collaboration between data scientists, domain experts, and healthcare professionals. Each group plays an essential role in ensuring that XAI models are not only accurate but also interpretable and actionable in real-world clinical settings.

Data scientists lead the development and refinement of XAI models, ensuring they can produce explanations that are understandable by non-technical stakeholders while maintaining the precision needed for diagnostics. This is no easy task, as XAI models must balance complexity with interpretability. It is very crucial to provide transparency, accountability, and insights into the behavior of XAI systems while addressing demographic inequalities and ensuring fairness in decision-making. By elucidating the factors influencing XAI outcomes, potential biases can be identified that may disproportionately affect certain demographic groups. Techniques like feature importance analysis allow us to see which variables are driving decisions, making it easier to spot and mitigate biases correctly. By leveraging XAI, models can be designed to adapt to different demographic contexts, which will help to avoid reinforcing historical inequalities, ensuring that all groups are treated fairly. However, addressing demographic inequalities or ensuring fairness in decision-making is a challenge for XAI classifiers, which can be minimized by the careful selection of bias and other controlling parameters.

Domain experts, including medical professionals, bring critical insights into the unique requirements and challenges of medical diagnostics. Their involvement is vital for determining which aspects of the data should be prioritized during the development of XAI systems. The input of these experts ensures that models are tailored to the specific needs of various medical fields. For example, the features that are most predictive in ophthalmology differ markedly from those in cardiology or nephrology. Domain-specific knowledge is necessary for ensuring that XAI systems provide clinically relevant explanations, especially in cases where understanding the reasoning behind an AI's decision can affect patient care.

Healthcare professionals, as the end users of these systems, play a vital role in validating the explanations provided by XAI models. They ensure that the AI outputs align with clinical knowledge and practice, helping to build trust in these technologies. Without validation, the use of XAI in critical medical diagnostics could introduce risks, as explanations that make sense from a technical standpoint might not be clinically relevant. Therefore, the collaboration between healthcare providers and data scientists is essential for improving both the accuracy and interpretability of AI-driven diagnostics.

One of the core challenges in developing XAI models is selecting the most appropriate method for generating explanations. Some XAI techniques prioritize feature importance to show which data points are most influential in making a decision, while others provide more abstract, human-readable explanations. The right choice of method depends on the clinical context and the specific needs of stakeholders, including healthcare providers and patients.

In addition to the challenge of selecting the right XAI methods, the complexity of healthcare data further complicates the development process. Electronic Health Records (EHRs) contain diverse information, including medical imaging, lab results, and historical data. This data is often incomplete or noisy, making it difficult to integrate and interpret. Data scientists must develop sophisticated models capable of processing these varied inputs, while still ensuring that their outputs can be understood by medical professionals.

Furthermore, overfitting remains a persistent issue in AI model development. Although models may perform well on training data, they often struggle when applied to more diverse patient populations. Overfitting not only reduces the reliability of predictions but also complicates the generation of trustworthy explanations across different healthcare settings. Addressing this problem requires rigorous model evaluation and careful tuning of parameters.

Another critical issue is managing the diverse needs of various stakeholders in healthcare. Patients often seek clear, simple explanations for diagnoses, while physicians require more detailed, technical insights. Meanwhile, hospital administrators may focus on operational efficiency and cost-effectiveness. XAI systems must be designed to meet these varied needs, which is a complex challenge requiring thoughtful design.

The successful deployment of XAI in healthcare hinges on a combination of technical innovation, domain-specific expertise, and collaboration across multiple disciplines. By addressing the challenges of data complexity, model interpretability, and stakeholder diversity, XAI systems can become robust tools for enhancing medical diagnostics, improving patient outcomes, and ensuring safer healthcare practices.

## 5.3. Future direction of research

### 5.3.1 Human-XAI Interaction

Motivated by the need for interpretability and transparency, XAI offers several opportunities for the adoption of AI in healthcare settings. However, it comes with its own risks, and inclusivity is needed to accommodate diverse stakeholders. Our review of existing literature has shown that user interaction with explanation interfaces and the XAI pipeline has not been thoroughly investigated. Hence, this area presents a great research opportunity. For example, some works developed guides for customizing explanation interfaces in user-focused ways to suit various users and application scenarios [233]. Another way of ensuring inclusivity is to consider exploring the human-computer interaction paradigm in designing XAI interfaces. In this context, an interdisciplinary collaboration between XAI and HCI presents a great opportunity. This could allow for more interactive visualizations and explanations using natural language.

### 5.3.2 Cost analysis and mitigation

AI systems are generally expensive to develop and deploy. These costs include labor, computation, data collection, and legal fees. While much work is ongoing on improving explanations in AI systems, not much is visible in studying the costs associated with XAI methods and ways of minimizing them. Even when comparing XAI methods, most works do not consider the cost implication of the various methods. Moreover, we understand that computational cost is a major concern for some XAI methods, such as counterfactual explanations. Hence, this also presents an important research direction. We believe cost analysis implications must be incorporated to develop or compare XAI methods in future works. This is key as it could guide entities on the implications of adopting XAI techniques in clinical settings while allowing for cost-benefit analysis. Another issue closely related to this is reducing computational costs. This area must be addressed effectively for some XAI techniques to become mainstream.

### 5.3.3 Regulatory Framework and Ethics

Integrating AI into critical healthcare operations raises considerable ethical questions and fairness challenges, a web of questions that XAI tries to answer to close the gap between humans and AI methods. However, XAI also has its binding moral and equity commitments. As the regulatory framework evolves to ensure the moral adoption of XAI methods, it is becoming challenging to protect patient rights, ensure data privacy, and preserve the integrity of medical practice. This is further exasperated by the rapid evolution of AI techniques, which could outpace the development of comprehensive ethical frameworks. Hence, regulation needs to balance between moral requirements and the need for innovation. Any developed framework must ensure transparency, fairness, and human-centricity. This also presents some research opportunities. For example, there is a need to understand how to incorporate regulatory restrictions into XAI systems to ensure that they give actionable outcomes. There is also the need to develop architecture and frameworks that ensure that XAI ensures that the architecture of any XAI system complies with regulatory requirements. Lastly, there is the need for a standardized XAI regulatory and ethics framework.

### 5.3.4 Choice of XAI methods and the effect of Noise

Different XAI has been employed to improve trust in ML outcomes and decisions. However, studies are lacking in selecting the most appropriate XAI methods for different ML techniques or specific healthcare applications. A framework for evaluating the effectiveness of XAI methods, considering the effects of noise and uncertainties, is also needed. The concept of uncertainty in the XAI domain needs to be formally defined and consolidated.

LRP has the capability to perform noise checks. LRP, for instance, works by attributing the output of a neural network back to its input features, thereby identifying which features are most relevant to the model's predictions. This method can help ensure that the features highlighted by the model are genuinely important, even in the presence of noise. On the other hand, some XAI methods rely directly on the source data to determine which features are important without explicitly accounting for noise. These methods may assume that the data is clean and noise-free, which is often not the case in real-world scenarios. As a result, these methods might highlight features that are not truly relevant, leading to erroneous interpretations of the model's decision-making process.

To address these issues, future research should focus on validating XAI models in noisy environments to minimize subjectivity in ML decisions. This involves rigorously testing XAI methods with data that include varying noise levels to see how well they can identify the truly important features. By doing so, researchers can ensure that the XAI methods used in ML-based diagnostic models are robust and reliable, thereby reducing the risk of misdiagnosis and improving patient safety.

Moreover, it is essential to develop standardized benchmarks and evaluation protocols for assessing the performance of XAI methods. These benchmarks should include datasets with known noise levels and well-defined important features. By comparing the performance of different XAI methods against these benchmarks, researchers can better understand each method's strengths and weaknesses.

Future work should also explore integrating multiple XAI techniques for improved performance. Combining the insights from different methods may make achieving a more comprehensive understanding of the model's decision-making process possible. For instance, using LRP in conjunction with other methods like SHAP or Grad-CAM could provide a more nuanced view of which features are most important for the model's predictions. Another important consideration is the transparency and interpretability of the XAI methods themselves. Users, particularly those in the medical field, need to understand how these methods work and how they derive their explanations. Providing clear and detailed documentation, along with visual aids, can help demystify these methods and make them more accessible to non-experts.

**5.3.5 Data unification and needs**

Medical information systems are also heterogeneous, which further complicates medical data and often makes interpretation challenging for various users. Considering the prospects of XAI in the healthcare sector, it is important to move towards Findable, Accessible, Interoperable, and Reusable (FAIR) data for both humans and machines. This would ensure interoperability and seamless integration of XAI across multiple platforms. A FAIR-based framework is necessary for the storage of actionable data.

It has been noted that it is critical to examine the XAI techniques for speed and resource consumption, particularly in healthcare settings with limited resources. However, in order to support clinical understanding, our approach places a higher priority on the level of explanation offered by various XAI methodologies. The significant trade-off between explainability, rapidity, and accuracy will surely be the subject of future research. We think this is an important subject for future research, and the AI community is actively working to find solutions to this problem.

**5.3.6 Domain knowledge integration and formal framework**

There is a need to consolidate sets of domain-specific basic principles for integration into XAI models. These could be made available as open source, and developers could pre-load them to their models based on their intended use. For example, publicly available and open basic knowledge and setup information in oncology should be integrated into any XAI pipeline meant for oncology-based healthcare. This should be the case for other specializations. In this light, a framework consolidating expert knowledge from various domains is necessary to formalize XAI practice.

**5.3.7  Unified XAI framework**

Despite efforts to formalize the XAI framework, there is a need for a unified XAI framework. The various stakeholders, such as patients, doctors, hospital authorities, pharmacies, diagnostic centers, and research centers, can be more easily accommodated via such a framework that ensures inclusivity and management of all the stakeholders involved. This framework will also ensure that XAI-based healthcare management is accessible to more people while ensuring accountability and trust.

Overall, the domain is just growing, and there are many open questions begging for answers. Hence, the field presents a vast opportunity for research and collaboration among the various stakeholders.

**5.3.7  The balance between accuracy and interpretability**

Since balancing accuracy with interpretability impacts the cost of developing and implementing XAI in healthcare and other domains, it is also an area needing research efforts. Moreover, computational limitations can have far reaching consequences in time-sensitive ailments demanding real-time decisions. Finding answers to the questions here will require consolidation of works done on AI methods to improve accuracy and the efforts on XAI techniques to improve interpretability at a reasonable computational burden. Overall, the domain is just growing, and there are many open questions begging for answers. Hence, the field presents a vast opportunity for research and collaboration among the various stakeholders.

*5.4 Integration of Explainable AI with Real-Time Health Monitoring Systems and Electronic Health Records (EHRs)*

**5.4.1 Data Flow in Real-Time Health Monitoring Systems**

Future research in real-time health monitoring systems must also emphasize the integration of Explainable AI (XAI) with Electronic Health Records (EHRs) to enhance the overall effectiveness of healthcare delivery. Accurate data collection from wearable devices, sensors, and mobile health apps, combined with advanced preprocessing methods, ensures the reliability of the data that flows into AI models. By connecting this continuous stream of health data to EHRs, healthcare practitioners gain real-time access to both historical and current patient information during consultations or emergencies.

The automated transmission of monitoring data to EHRs reduces errors and ensures that patient records are up to date, enabling providers to track health trends over time. This integration supports XAI by giving clinicians a comprehensive view of the patient's condition, combining past data with real-time insights to create detailed patient profiles. These profiles allow for more personalized and informed treatment plans. The transparency offered by XAI, along with the seamless flow of data into EHRs, further ensures that healthcare professionals can understand the reasoning behind AI-generated predictions and recommendations, improving decision-making and patient outcomes.

**5.4.2 Potential Delays and Mitigation Strategies**

A fundamental difficulty in real-time health monitoring systems is reducing data processing and transmission delays, which might impede prompt responses. Latency can be caused by a variety of variables, including limited network bandwidth, processing time inside AI models, and integration delays with EHR systems. To solve these issues, numerous techniques are used:
- Edge Computing: Data pretreatment and early analysis can take place closer to the data source (for example, on a wearable device or a local gateway), minimizing the amount of data transferred to central servers and lowering latency.
- Efficient Algorithms: The AI models employed in real-time health monitoring are designed for speed while maintaining accuracy. Lightweight models can handle data quickly, resulting in real-time responsiveness.
- Redundant Data Streams: Multiple data transmission paths (such as Wi-Fi and cellular networks) ensure that data flows continuously even if one of them is delayed or interrupted.

**5.4.3 Adaptability of the System**

The AI system's versatility is critical for serving varied patient groups and meeting changing healthcare demands. The system is meant to be flexible, so it can:
- Personalize Monitoring: Tailor monitoring and prediction algorithms to each patient's specific health conditions, risk factors, and medical history. For example, a patient with diabetes may have a different monitoring strategy than one with cardiovascular disease.
- Evolving Data Patterns: The system is constantly learning and adapting depending on new data, ensuring that it remains relevant and accurate even as patient conditions alter over time.

Integration of Various Health Data Sources: The system is interoperable with a wide range of health equipment and apps, making it suitable for a variety of healthcare settings and medical practices.

**6. Conclusion**

In conclusion, this research paper has extensively explored the application and impact of XAI in human health monitoring, particularly within Gait, Parkinson's, stroke, depression, cancer, Heart Disease, Alzheimer's Disease, and COVID-19. The findings from the reviewed papers clearly demonstrate that XAI not only enhances the accuracy and efficiency of human health monitoring but also significantly improves the transparency and interpretability of the diagnostic process. Moreover, the reported patient outcomes and satisfaction levels underscore the practical benefits of integrating XAI into clinical practice. By providing tailored, understandable explanations of AI-driven decisions, XAI fosters patient trust and facilitates more informed and collaborative decision-making processes between patients and healthcare providers. Looking ahead, the continued development of XAI models promises further improvements in the personalized and precision medicine landscape. The potential for these technologies to process complex datasets efficiently, adapt to individual patient needs dynamically, and offer insights into treatment efficacy is immense. However, it is crucial to continuously address the challenges and limitations identified, such as the complexity-explainability trade-off, data representation, computational constraints, and ethical considerations. The integration of XAI into healthcare is a testament to the transformative power of AI in enhancing patient care and treatment outcomes. As AI continues to evolve, it holds the potential to revolutionize various aspects of healthcare, making diagnostics and treatment more accurate, efficient, and patient-centric. This research underscores the importance of ongoing innovation and interdisciplinary collaboration in harnessing the full potential of XAI in healthcare, paving the way for a future where technology and medicine converge to improve human health and well-being.

## 7. Funding

The authors extend their appreciation to the Deanship of Research and Graduate Studies at King Khalid University for funding this study through Large Research Project under grant number RGP2/254/45.